\documentclass[aps,prd,onecolumn,showpacs,amssymb]{revtex4}
\usepackage{graphicx,bm,color}
\usepackage{amsmath}
\usepackage{amssymb}
\usepackage{amsfonts}
\newcommand{\be}{\begin{equation}}
\newcommand{\ee}{\end{equation}}
\newcommand{\bea}{\begin{eqnarray}}
\newcommand{\eea}{\end{eqnarray}}
\newcommand{\beaa}{\begin{eqnarray*}}
\newcommand{\eeaa}{\end{eqnarray*}}

\newcommand{\nn}{\nonumber \\}

\allowdisplaybreaks[4]

\begin{document}

\title{Oscillating universe in massive bigravity}
\author{ M. Mousavi}\email{ mousavi@azaruniv.ac.ir}
\author{F. Darabi}\email{ f.darabi@azaruniv.ac.ir}

\affiliation{Department of Physics, Azarbaijan Shahid Madani University, Tabriz, 53714-161 Iran
\\ Iran National Science Foundation: INFS, Tehran, 1439634665 Iran}

\begin{abstract}
In this paper, in the framework of massive bigravity, we study all possible cosmic evolutions by using a method in which the modified Friedmann equation is written in a form where the scale factor evolves like the motion of a particle under a ``potential". Massive bigravity provides this potential with the most general mass interaction term which can create new circumstances to find different kinds of cosmological evolutions in the early universe. We classify all possible cosmic evolutions according to the classifications of the energy density as dust, radiation and dust with phantom. Oscillating universe and Einstein static state which exist initially may show a useful property of early universe, obtained in this model, in which the initial
singularity is avoided. Bouncing universe extracted in the massive bigravity model can present a reasonable cosmic evolutionary behavior having a big bang initial point with expansion phase and switching to contraction phase
leading  to final big crunch point. The large-valued graviton mass $m$ in the early times causes a very small  $a_{\rm{S}}$ (The Einstein static state scale factor) and $\lambda=\rho_{0}a_{0}^{3}$ a constant parameter constructed of the present day energy density and scale factor, respectively.

\end{abstract}


\maketitle

\section{Introduction \label{Sec1}}
The discovery of current cosmic acceleration by means of the type Ia supernovas data \cite{1,2}, the cosmic microwave background radiation \cite{3} and the large scale structure \cite{4,5}, has established a renewed interest in theories that modify standard gravity. Some people believe that the theory of general relativity is no longer valid on the cosmological scale and needs a modification \cite{4d,5d}. Therefore, many modified theories have been proposed to explain the present accelerating expanding universe without the need for an unknown dark energy element. On
the other hand, giving a tiny mass to graviton is one way to modify the general relativity which can give rise to a small cosmological term leading to currently observed accelerated expanding universe. The old history of massive gravity dates back to 1939, when Fierz and Pauli published their linear model \cite{6}. Nevertheless, the Fierz-Pauli action dose not describe the linearized Einstein gravity in the zero-mass limit and also cannot satisfy the solar system tests due to the van Dam-Veltman-Zakharov discontinuity \cite{7}. The Vainshtein mechanism can avoid this discontinuity by introducing nonlinear interactions  \cite{8}. Consequently, the nonlinear terms yield a ghost called the Boulware-Deser ghost \cite{9,10}. Eventually, a ghost-free nonlinear massive gravity theory was established successfully by de Rham, Gabadadze and Tolly \cite{11}.
In \cite{12} one can review the steps leading to the modern approach.

 In these new forms of massive gravity, beside the metric a second tensor field plays a key role. This theory of massive gravity was later shown to be ghost-free \cite{13}. People have  shown that the massive gravity cosmological model does not yield the flat Friedmann-Robertson-Walker universe \cite{14,57,58}, however this result is not supported by the recent observational results.

  The authors of references \cite{15,16} make the second tensor field dynamical, just as the standard metric, although only the latter is coupled to matter. As a result, we face with the massive bigravity  in which the theory is kept ghost-free,  allowing cosmologically viable solutions, and also dose not have the previous cosmological problems mentioned in massive gravity. This modified model involves two dynamical metrics in a completely symmetric way that obviates the aether-like concept of reference metric in massive gravity. In massive bigravity model, some cosmological solutions have been derived in \cite{17,18,188,190,191}, some other solutions  were studied via cosmological perturbation \cite{192,193} and additionally, the Einstein static universe (ES) in this massive bigravity theory was studied in \cite{19} by the authors in which we found that there exist stable ES solutions which can avoid the big bang singularity.

  This line of investigation has motivated us to go through seeking the deferent types of cosmological evolution  at early universe in the massive bigravity model as has been done in the massive gravity model, DGP braneworld scenario and Horova-Lifshitz gravity \cite{20,21,22}. These works  follow a simple method in which one write the Friedmann equation in a form such that the scale factor evolution behaves like the motion of a particle in a potential,
and this makes it possible to investigate cosmic evolutions like oscillating universe, bouncing or the ES universe. In this work, we are going to apply this method to study all possible cosmic evolutions in massive bigravity. The organization of this paper is as follows. Section II is devoted to the modification of the Friedmann equations in massive bigravity model and definition of all possible cosmological evolution types. In sections III and IV, we derive the details of the ES solution and classify the resulted cosmic evolutions with their extracted conditions for matter-dominated and radiation-dominated universes, respectively. In section V, we investigate the cosmic evolutions in a universe with dust and phantom and finally, we present the conclusion part in section VI.

\section{THE FRIEDMANN EQUATIONS IN MASSIVE BIGRAVITY \label{Sec2}}

  Massive bigravity is introduced by the action \cite{14}

\be
\label{Fbi1}
S_{{\rm bi}}=-\frac{M_{g}^{2}}{2}\int d^{4}x\sqrt{- {\rm det} g}R-\frac{M_{f}^{2}}{2}\int d^{4}x\sqrt{- {\rm det} f}\tilde{R}+
m^{2}M_{g}^{2}\int d^{4}x\sqrt{- {\rm det} g}\sum_{n=0}^{4}\beta_{n}e_{n}\left(\sqrt{g^{-1}f}\right)+\int d^{4}x \sqrt{- {\rm det} g}~\mathcal{L}_{m},
\ee
where $g_{\mu\nu}$, $f_{\mu\nu}$ are two dynamical metrics with corresponding Ricci scalars $R$, $\tilde{R}$, respectively,  ${M}_{g}$ and  ${M}_{f}$ are  two Planck mass scales for  $g_{\mu\nu}$ and $f_{\mu\nu}$ respectively,
  $\mathcal{L}_{m}\equiv\mathcal{L}_{m}~(g,\Phi)$ is the matter Lagrangian containing an scalar field $\Phi$, the parameter $m$ describes the mass of graviton or the massive spin-2 field, and $\beta_{n}$ are some parameters
of the model.
The square root matrix $\sqrt{g^{-1}f}$ is defined by $\left(\sqrt{g^{-1}f}\right)^{\mu}~_{\rho}\left(\sqrt{g^{-1}f}\right)^{\rho}~_{\nu}=g^{\mu\rho}f_{\rho\nu}=X^{\mu}~_{\nu}$. For the trace of this  tensor or general matrix as $X^{\mu}~_{\mu}$ or  $[X]$, $e_{n}(X)$'s  are elementary symmetric polynomials of the eigenvalues of $X$:
\begin{align}\label{Fbi2}
e_{0}(X)=&1,~~e_{1}(X)=[X],~~e_{2}(X)=\frac{1}{2}\left([X]^{2}-[X^{2}]\right),\nn
e_{3}(X)=&\frac{1}{6}\left([X]^{3}-3[X][X^{2}]+2[X^{3}]\right),\nn
e_{4}(X)=&\frac{1}{24}\left([X]^{4}-6[X]^{2}[X^{2}]+3[X^{2}]^{2}+8[X][X^{3}]-6[X^{4}]\right),\nn
e_{i}(X)=&0~~{\rm for}~~i>4.
\end{align}
According to a nonlinear ADM analysis of Hassan and Rosen in \cite{23}, the action (\ref{Fbi1}) is explicitly ghost-free and describes 7 propagating degrees of freedom. Ignoring the matter coupling part, the action is invariant under the following
exchanges,

\begin{align}\label{Fbi3}
g\leftrightarrow f,~~~~~\beta_n \rightarrow \beta_4-n,~~~~~M_g \leftrightarrow M_f,~~~~m^2\rightarrow m^2M_g^2/M_f^2.
\end{align}
It is noticeable that setting $\beta_3=0$ in (\ref{Fbi1}) eliminates the highest order interaction term in $\sqrt{g^{-1}f}$. However, according to (\ref{Fbi3}) in $\sqrt{f^{-1}g}$ we still have a cubic order interaction term which can in turn be eliminated by setting  $\beta_1=0$. Eventually, setting $\beta_1=\beta_3=0$ leads to the "minimal" massive bigravity action, which is the simplest in the class.

 Now, we consider the variation of the action (\ref{Fbi1})
with respect to $g_{\mu\nu}$ and $f_{\mu\nu}$, respectively as
\begin{align}
\label{Fbi4}
0=R_{\mu\nu}-\frac{1}{2}g_{\mu\nu}R+\frac{m^{2}}{2}\sum_{n=0}^{3}(-1)^n \beta_{n} \left[g_{\mu\lambda}Y_{(n)\nu}^{\lambda}\left(\sqrt{g^{-1}f}\right)+g_{\nu\lambda}Y_{(n)\mu}^{\lambda}\left(\sqrt{g^{-1}f}\right)\right]
-\frac{T_{\mu\nu}}{M_{g}^2},
\end{align}
and
\begin{align}\label{Fbi5}
0=\tilde{R}_{\mu\nu}-\frac{1}{2}f_{\mu\nu}\tilde{R}+\frac{m^{2}}{2M_{*}^2}\sum_{n=0}^{3}(-1)^n \beta_{4-n} \left[f_{\mu\lambda}Y_{(n)\nu}^{\lambda}\left(\sqrt{f^{-1}g}\right)+f_{\nu\lambda}Y_{(n)\mu}^{\lambda}\left(\sqrt{f^{-1}g}\right)\right],
\end{align}
where
\begin{align}\label{Fbi6}
M_{*}^{2}\equiv \frac{M_f^2}{M_g^2}.
\end{align}
Additionally, the matrices $Y_{(n)\mu}^{\lambda}\left(X\right)$ introduced in the above field equations are given by
\begin{align}\label{Fbi7}
Y_{(0)}\left(X\right)=&1,~~~Y_{(1)}\left(X\right)=X-1\left[X\right],\nn Y_{(2)}\left(X\right)=&X^2-X\left[X\right]+\frac{1}{2}1\left(\left[X\right]^2-\left[X^2\right]\right),\nn
Y_{(2)}\left(X\right)=&X^3-X^2\left[X\right]+\frac{1}{2}X\left(\left[X\right]^2-\left[X^2\right]\right)-\frac{1}{6}1\left(\left[X\right]^3-
3\left[X\right]\left[X^2\right]+2\left[X^3\right]\right).
\end{align}
The covariant conservation of $T_{\mu\nu}$ beside the field equation (\ref{Fbi4}) leads to the Bianchi constraint for the metric $g_{\mu\nu}$
\be\label{Fbi8}
0=\nabla^{\mu} \sum_{n=0}^{3}\left(-1\right)^{n}\beta_{n} \left[g_{\mu\lambda}Y_{(n)\nu}^{\lambda}\left(\sqrt{g^{-1}f}\right)+g_{\nu\lambda}Y_{(n)\mu}^{\lambda}\left(\sqrt{g^{-1}f}\right)\right].
\ee
The field equation (\ref{Fbi5}) also gives us the Bianchi constraint corresponding to the metric $f_{\mu\nu}$ as
\be\label{Fbi9}
0=\tilde{\nabla}^{\mu} \sum_{n=0}^{3}\left(-1\right)^{n}\beta_{4-n} \left[f_{\mu\lambda}Y_{(n)\nu}^{\lambda}\left(\sqrt{f^{-1}g}\right)+f_{\nu\lambda}Y_{(n)\mu}^{\lambda}\left(\sqrt{f^{-1}g}\right)\right],
\ee
where $\tilde{\nabla}^{\mu}$ implies the covariant derivatives with respect to the metric $f_{\mu\nu}$. We can realize that two above Bianchi constraints are equivalent as a result  of  invariance of the interaction term under the general coordinate transformations of  two metrics, so we just use the constraint (\ref{Fbi8}).
We consider a homogeneous and isotropic Friedmann-Robertson-Walker (FRW) universe with three-dimensional spatial curvature $\kappa=\pm1$  for both metrics
\be
\label{Fbi10}
ds_{g}^{2}=-dt^{2}+a(t)^{2}\left(\frac{dr^{2}}{1-\kappa r^{2}}+r^{2}d\theta^{2}+r^{2}\sin^{2}\theta d\varphi^{2}\right),
\ee
\be
\label{Fbi11}
ds_{f}^{2}=-c(t)^{2}dt^{2}+b(t)^{2}\left(\frac{dr^{2}}{1-\kappa r^{2}}+r^{2}d\theta^{2}+r^{2}\sin^{2}\theta d\varphi^{2}\right),
\ee
where $a(t)$ is the cosmic scale factor related to $g_{\mu\nu}$ and $b(t)$ is the one related to $f_{\mu\nu}$. Obviously, $c(t)$ (the lapse function of $f_{\mu\nu}$ metric) is a function of time and note that we do not have any more freedom to choose $c=1$ nor $c=b$.

For the metrics (\ref{Fbi10}) and (\ref{Fbi11}), the Bianchi constraint (\ref{Fbi8}) reduces to

\be\label{Fbi12}
\frac{3m^2}{a} \left(\beta_{1}+2\gamma \beta_{2}+\gamma^{3}\beta_{3}\right)\left(\dot{b}-\dot{a}c\right)=0,
\ee

where  $\gamma\equiv\frac{b(t)}{a(t)}$. If the first parenthesis vanishes, we will find solutions with $b\propto a$ which leads us to the ordinary GR equations including a cosmological constant of order $m^2$, independent of any dynamics of $f_{\mu\nu}$. Other than this choice, we can consider the vanishing of
the second parenthesis
which leads to

\be\label{Fbi13}
c(t)=\frac{\dot{b}}{\dot{a}}~.
\ee

Let us now consider the  source structure. As specified in (\ref{Fbi1}) the matter source is just coupled to the metric $g_{\mu\nu}$, so by assuming an equation of state of the normal form $P(t)=\omega \rho(t)$ in the minimal coupling of the matter to gravity $g_{\mu\nu}$, and defining $\lambda=\rho_{0}a_{0}^{3}$ ($\lambda$ is a positive constant), we have $\rho=\frac{\lambda}{a^{3\left(1+\omega\right)}}$.

In the following, we take $\omega=0$ or $\frac{1}{3}$, which corresponds to pressureless matter or radiation-dominated universe, respectively.
 In spite of the fact that massive gravity and also massive bigravity can explain the present accelerated cosmic expansion, they may also play an important role in the very early universe with very small scale factor. In the present paper, we follow the main idea of the work \cite{22} in massive gravity theory which have been done for a spatially flat universe with a positive constant vacuum energy $\rho$. Accordingly, we  plan to study all the possible cosmic evolutions of the early universe in massive bigravity. It is worth  mentioning
that we have already studied the static cosmological solutions and their stability at background level in the framework of massive bigravity theory with FRW metrics which led to a class of new solutions interpreted as the Einstein static universe \cite{19}. In that paper,  we have shown that the non-vanishing size of initial scale factor of Einstein static universe which depends on the non-vanishing spatial curvature of FRW metrics and  the graviton's mass, can resolve the big bang singularity.

Defining $\widetilde{\rho}\equiv \frac{\rho}{3M_{g}^{2}m^{2}}$ we write the Friedmann equations corresponding to $g_{\mu\nu}$ and the combination of the $g_{\mu\nu}$-$f_{\mu\nu}$  respectively as

\be\label{Fbi14}
\frac{H^{2}}{m^{2}}+\frac{\kappa}{m^{2}a^{2}}=\frac{\beta_{3}}{3}\gamma^{3}+\beta_{2}\gamma^{2}+\beta_{1}\gamma+\frac{\beta_{0}}{3}+\widetilde{\rho},
\ee

\be\label{Fbi15}
\frac{\beta_{3}}{3}\gamma^{4}+\left(\beta_{2}-\frac{\beta_{4}}{3M_{*}^{2}}\right)\gamma^{3}+\left(\beta_{1}-\frac{\beta_{3}}{M_{*}^{2}}\right)\gamma^{2}
+\left(\tilde{\rho}+\frac{\beta_{0}}{3}-\frac{\beta_{2}}{M_{*}^{2}}\right)\gamma-\frac{\beta_{1}}{3M_{*}^{2}}=0,
\ee
where $H=\frac{\dot{a}}{a}$ is the Hubble parameter of the scale factor $a(t)$ and $t$ is the cosmic time.

Actually, in order to find the behavior of the scale factor $a$  we should focus on both equations (\ref{Fbi14}) and (\ref{Fbi15}), because the equation
(\ref{Fbi15}) gives $\gamma$ merely in terms of the scale factor $a$,  and
using the definition of $\gamma(a)=b/a$ and the equation (\ref{Fbi14}) we can find the time evolution of $b(t)$ beside the time evolution of $a(t)$.

 As a result, We rewrite the Friedmann equation of $g_{\mu\nu}$ (\ref{Fbi14}) in the following form

\be\label{Fbi16}
\dot{a}^2+V(a)=0,
\ee
where
\be\label{Fbi17}
V(a)=\kappa-m^{2}a^{2}\left(\frac{\beta_{3}}{3}\gamma^{3}+\beta_{2}\gamma^{2}+\beta_{1}\gamma+\frac{\beta_{0}}{3}+\widetilde{\rho}\right).
\ee

Therefore, we can regard $V(a)$ as a potential and the scale factor $a$ as that of a particle moving in a potential $V$ and clearly, this potential must satisfy the condition $V(a)\leq 0$, resulted from equation (\ref{Fbi16}), which gives the meaningful ranges of $a$ as the universe evolves. Thus, we can classify the types of  universe by the sign of $\kappa$ and  the values of parameters $\lambda$, $\omega$ and $\beta_{i}$.

All types of  universe in massive bigravity theory are categorized as follows:
\begin{itemize}
\item(1) \emph{Bounce}  \\For $a\in [a_{T},\infty)$, if the potential gets negative values ($V(a)\leq0$) and the equality holds at $a=a_{T}$, the universe initially contracts from an infinite scale factor. Eventually, it turns back at a finite scale factor $a_{T}$ and then expands to infinity forever.
\item(2) \emph{Oscillation} \\For $a\in [a_{\rm{min}},a_{\rm{max}}]$, $V(a)\leq0$ and the equality occurs at $a=a_{\rm{min}}$ and $a=a_{\rm{max}}$. Therefore, the universe oscillates between two finite scale factors.
\item(3) \emph{BB}$\Rightarrow $\emph{BC}\\ For $a\in (0,a_{T}]$, $V(a)\leq0$ and the equality holds at $a=a_{T}$. The universe  starts from a big bang (BB) and expands. Finally, it turns back at $a=a_{T}$ and contracts to a big crunch (BC). Note that, $a_{T}$ is the scale factor where the universe turns back from expansion to contraction.
\item(4) \emph{BB}$\Rightarrow \infty$ or $\infty\Rightarrow$\emph{BC}\\
For $a>0$ we have $V(a)<0$. The universe starts from a big bang and expands forever, or the universe always contracts to a big crunch.
\end{itemize}

\section{THE EVOLUTION OF A MATTER-DOMINATED UNIVERSE IN THE MASSIVE BIGRAVITY MODEL}

When the universe is dominated by pressureless matter with $\omega=0$, the cosmic energy density can be expressed as $\rho=\frac{\lambda}{a^{3}}$. As a result, the potential becomes

\be\label{Fbi18}
V(a)=\kappa-m^{2}a^{2}\left(\frac{\beta_{3}}{3}\gamma^{3}+\beta_{2}\gamma^{2}+\beta_{1}\gamma+\frac{\beta_{0}}{3}+\frac{\lambda}{3M_{g}^{2}m^{2}a^{3}}
\right).
\ee

Since, $\rho$ is a function of scale factor, according to (\ref{Fbi15}) we obtain a relation between $\gamma$ and $a$.
Considering (\ref{Fbi14}), $\gamma$ appears cubic so it will be obvious that we have terms with deferent powers of scale factor which definitely makes the potential term too complicated to be explicitly solved. Looking for simplicity, we go through another interesting class of cosmological evolutions subject
to $\beta_{1}=\beta_{3}=0$.
 Setting $\beta_{3}=0$ in the action (\ref{Fbi1}) causes the highest order interaction term in $\sqrt{g^{-1}f}$ to be eliminated. Nevertheless, looking at (\ref{Fbi3}) we still have a cubic order interaction term in $\sqrt{f^{-1}g}$ which can be eliminated by choosing $\beta_{1}=0$. In this context, choosing $\beta_{1}=\beta_{3}=0$ leads to the ``minimal" massive bigravity action where the interaction terms are of the lowest order  for both $\sqrt{g^{-1}f}$ and $\sqrt{f^{-1}g}$. This leaves only nonlinear interactions with quadratic order in both sectors and the action (\ref{Fbi1}) looks like a nonlinear one with a mass potential.
On the other hand, at the early universe with a large energy density $\tilde{\rho}$, we will be sure that the term $\left(\tilde{\rho}+\frac{\beta_{0}}{3}-\frac{\beta_{2}}{M_{*}^{2}}\right)$ becomes positive and in order  to have a real-valued $\gamma$, one should have $\beta_{2}<\frac{\beta_{4}}{3M_{*}^{2}}$.

With these arrangements, equations (\ref{Fbi14}) and (\ref{Fbi15}) are reduced to

\be\label{Fbi19}
\frac{H^{2}}{m^{2}}+\frac{\kappa}{m^{2}a^{2}}=\beta_{2}\gamma^{2}+\frac{\beta_{0}}{3}+\widetilde{\rho},
\ee

\be\label{Fbi20}
\left(\beta_{2}-\frac{\beta_{4}}{3M_{*}^{2}}\right)\gamma^{2}
+\left(\tilde{\rho}+\frac{\beta_{0}}{3}-\frac{\beta_{2}}{M_{*}^{2}}\right)=0,
\ee

Using (\ref{Fbi19}) and (\ref{Fbi20}), we simply can eliminate $\gamma$ in the general potential $V(a)$, (\ref{Fbi18}) as follows

\be\label{Fbi21}
V(a)=\kappa+m^{2}a^{2}\times\frac{\frac{-3\beta_{2}^{2}}{\beta_{4}}+\frac{\beta_{0}}{3}+\frac{\lambda}{3a^{3\left(1+\omega\right)}m^{2}M_{g}^{2}}}{
\frac{3M_{*}^{2}\beta_{2}}{\beta_{4}}-1}.
\ee

  It is noticeable that, we are looking for a static universe with a minimal scale factor $a=a_{\rm{s}}>0$ which satisfies $V(a_{\rm{s}})=0$ and $V'(a_{\rm{s}})=0$ (implying that both  cosmic expansion speed and acceleration are equal to zero), so we apply  two above conditions to obtain ES quantities $a_{\rm{S}}$ and $\lambda$.\\

  \textbf{Case I:} $\kappa=1$\\

 Using (\ref{Fbi21}), we find the potential as
\be\label{Fbi22}
V(a)=\frac{a^2 \left(-\frac{3 \beta _2^2}{\beta _4}+\frac{\beta _0}{3}+\frac{\lambda }{3 a^3 m^2 M_g^2}\right) m^2}{\frac{3 \beta _2 M_*^2}{\beta _4}-1}+1.
\ee
 Differentiating $V(a)$ with respect to $a$, we obtain the following result
\be\label{Fbi23}
V'(a)=\frac{2 a m^2 \left(-\frac{3 \beta _2^2}{\beta _4}+\frac{\beta _0}{3}+\frac{\lambda }{3 a^3 m^2 M_g^2}\right)}{\frac{3 \beta _2 M_*^2}{\beta _4}-1}-\frac{\lambda
   }{a^2 M_g^2 \left(\frac{3 \beta _2 M_*^2}{\beta _4}-1\right)}.
\ee
Here, $V'(a)=\frac{dV(a)}{da}$.
To obtain an ES solution, by combining $V(a)=0$ and $V'(a)=0$ we find a relation between $\lambda$ and three other model parameters $\beta_{0}$, $\beta_{2}$ and $\beta_{4}$ as follows

\begin{equation}\label{Fbi24}
\lambda=\lambda^{\pm}=\pm \frac{2M_{g}^{2}\left(\beta_{4}-3\beta_{2}M_{*}^{2}\right)^{\frac{3}{2}}}{\beta_{4}\sqrt{m^{2}
\left(-9\beta_{2}^{2}+\beta_{0}\beta_{4}\right)}},
\end{equation}
and also we extract the following static state solutions

\be\label{Fbi25}
a_{\rm{s}}=a_{\rm{s}}^{\pm}=\pm \frac{\sqrt{\beta_{4}-3\beta_{2}M_{*}^{2}}}{\sqrt{m^{2}
\left(-9\beta_{2}^{2}+\beta_{0}\beta_{4}\right)}}.
\ee

Now, we insert $\lambda^{\pm}$ into $V(a)=0$ under the condition $V'(a)=0$, then we find $a_{\rm{T}}^{\pm}$

\be\label{Fbi26}
a_{\rm{T}}=a_{\rm{T}}^{\pm}=\frac{m^{4}\left(-9\beta_{2}^{2}+\beta_{0}\beta_{4}\right)^{2}\left(\beta_{4}-3\beta_{2}M_{*}^{2}\right)
+\left(\mp m^{6}\left(-9\beta_{2}^{2}+\beta_{0}\beta_{4}\right)^{3}\left(\beta_{4}-3\beta_{2}M_{*}^{2}\right)^{\frac{3}{2}}\right)^{\frac{2}{3}}}{
m^{3}\left(9\beta_{2}^{2}-\beta_{0}\beta_{4}\right)\sqrt{m^{2}\left(-9\beta_{2}^{2}+\beta_{0}\beta_{4}\right)}\left(\mp m^{6}\left(-9\beta_{2}^{2}+\beta_{0}\beta_{4}\right)^{3}\left(\beta_{4}-3\beta_{2}M_{*}^{2}\right)^{\frac{3}{2}}\right)^{\frac{1}{3}}}.
\ee
Note that if $a_{\rm{T}}$ gives positive value, it corresponds to bouncing or turning radius  of the universe (the radius where the universe bounces or turns around). As we said before, $\lambda$ and $a$ should be positive so we just consider $\lambda^{+}$, $a_{\rm{S}}^{+}$ and $a_{\rm{T}}^{+}$.

The potential can be rewritten as
\be\label{Fbi27}
V(a)=\frac{m^{2}\left(\frac{-3\beta_{2}^{2}}{\beta_{4}}+\frac{\beta_{0}}{3}\right)}{a\left(\frac{3M_{*}^{2}\beta_{2}}{\beta_{4}}-1\right)}\times
\left(a^{3}+\frac{\left(\frac{3M_{*}^{2}\beta_{2}}{\beta_{4}}-1\right)a}{m^{2}\left(\frac{-3\beta_{2}^{2}}{\beta_{4}}+\frac{\beta_{0}}{3}\right)}
+\frac{\lambda}{3m^{2}M_{g}^{2}\left(\frac{-3\beta_{2}^{2}}{\beta_{4}}+\frac{\beta_{0}}{3}\right)}\right),
\ee
which  gives a cubic equation if $V(a)=0$ . It is worth to say that when $\lambda$ takes deferent values, the number of real roots for the equation $V(a)=0$ becomes deferent. For instance, $\lambda=\lambda^{+}$ allows the existence of three real roots but two of them are double corresponding to an unstable ES solution. For more clearness, in what follow we divide our discussion into two classifications, ie., $0< \lambda \leq \lambda^{+}$ and $\lambda > \lambda^{+}$, respectively.\\\\

~~~~~~~~~~~~~~~~~~~~~~~~~~~~~~~~~~~~~~~~~~~~~~~~~~~~~~~~~~~\textbf{A}. $0<\lambda \leq \lambda^{+}$\\

For this case, $V(a)=0$ yields a cubic equation of $a$, which yields three real roots $a_{1}$, $a_{2}$ and $a_{3}$. Using these real roots, the equation (\ref{Fbi27}) can be written as

\be\label{Fbi28}
V(a)=\frac{m^{2}\left(\frac{-3\beta_{2}^{2}}{\beta_{4}}+\frac{\beta_{0}}{3}\right)}{a\left(\frac{3M_{*}^{2}\beta_{2}}{\beta_{4}}-1\right)}
\left(a-a_{\rm{T}}\right)\left(a-a_{\rm{min}}\right)\left(a-a_{\rm{max}}\right),
\ee
in which we have defined $a_{1}=a_{T}$, $a_{2}=a_{\rm{min}}$ and $a_{3}=a_{\rm{max}}$.\\

~~~~~~~~~~~~~~~~~~~~~~~~~~~~~~~~~~~~~~~~~~~~~~~~~~~~~~~~\textbf{1}.Three positive roots\\

By assuming $0\leq a_{\rm{T}}\leq a_{\rm{min}}\leq a_{\rm{max}} $, we have

\begin{align}\label{Fbi29}
V(a)=&
\frac{m^{2}\left(\frac{-3\beta_{2}^{2}}{\beta_{4}}+\frac{\beta_{0}}{3}\right)}{a\left(\frac{3M_{*}^{2}\beta_{2}}{\beta_{4}}-1\right)}
\left(a-a_{\rm{T}}\right)\left(a-a_{\rm{min}}\right)\left(a-a_{\rm{max}}\right)=
\frac{m^{2}\left(\frac{-3\beta_{2}^{2}}{\beta_{4}}+\frac{\beta_{0}}{3}\right)}{a\left(\frac{3M_{*}^{2}\beta_{2}}{\beta_{4}}-1\right)}\times \nn&
\left(a^{3}-\left(a_{\rm{min}}+a_{\rm{max}}+a_{\rm{T}}\right)a^{2}+\left(a_{\rm{min}}a_{\rm{max}}+a_{\rm{T}}a_{\rm{min}}+
a_{\rm{max}}a_{\rm{T}}\right)a-a_{\rm{min}}a_{\rm{max}}a_{\rm{T}}\right).
\end{align}
Comparing the above equation with the equation (\ref{Fbi27}) implies that

\begin{align}\label{Fbi30}
a_{\rm{min}}+a_{\rm{max}}+a_{\rm{T}}=0,
\end{align}

\begin{align}\label{Fbi31}
a_{\rm{min}}a_{\rm{max}}+a_{\rm{T}}a_{\rm{min}}+a_{\rm{max}}a_{\rm{T}}=\frac{\left(\frac{3M_{*}^{2}\beta_{2}}{\beta_{4}}-1\right)}{
m^{2}\left(\frac{-3\beta_{2}^{2}}{\beta_{4}}+\frac{\beta_{0}}{3}\right)},
\end{align}
and
\begin{align}\label{Fbi32}
a_{\rm{min}}a_{\rm{max}}a_{\rm{T}}=\frac{\lambda}{3m^{2}M_{g}^{2}\left(\frac{3\beta_{2}^{2}}{\beta_{4}}-\frac{\beta_{0}}{3}\right)}.
\end{align}

Before going through the discussion, we realize that using the terms $\lambda^{+}$ and $a_{\rm{s}}^{+}$ in (\ref{Fbi24}) and (\ref{Fbi25}), respectively,   a condition is imposed on the  free parameters
of the model
as \begin{align}\label{Fbi33}
\beta_{4}>\rm{Max} \left\{\frac{9\beta_{2}^{2}}{\beta_{0}},\,3\beta_{2}M_{*}^{2}\right\},
\end{align}
such that they become real-valued. $\beta_{0}$ and $\beta_{4}$ are playing the role of  cosmological constants of $g_{\mu\nu}$ and $f_{\mu\nu}$, respectively and thus we assume that they are positive. Figure 1. plotted for fixed $\beta_{0}=3$, shows the allowed range of free parameters $\beta_{2}$ and $\beta_{4}$ in which one can keep $\gamma$ and the quantities in equations (\ref{Fbi24})-(\ref{Fbi26}) real.
\begin{figure*}[ht]
  \centering
  \includegraphics[width=2.5in]{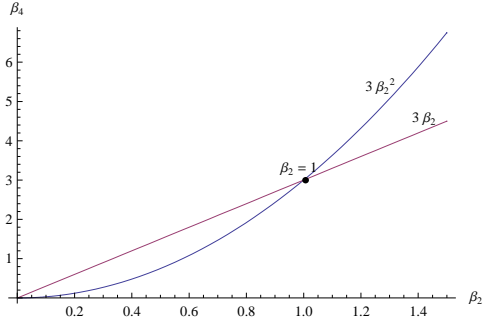}\hspace{2cm}
  \caption{Phase diagram of universes in $\left(\beta_{2},\beta_{4}\right)$ plane when $\gamma$, $\lambda^{+}$, $a_{\rm{S}}^{+}$ and $a_{\rm{T}}^{+}$ are real in region $\beta_{4}>\rm{Max} \{\frac{9\beta_{2}^{2}}{\beta_{0}},3\beta_{2}M_{*}^{2}\}$ with $M_{*}=1$.}
  \label{stable}
\end{figure*}
 Accordingly, the coefficient of $a^{3}$ term namely $\frac{\frac{-3\beta_{2}^{2}}{\beta_{4}}+\frac{\beta_{0}}{3}}{\frac{3M_{*}^{2}\beta_{2}}{\beta_{4}}-1}$ in the potential which plays a crucial role in determining the shape of the potential $V(a)$, becomes negative-valued. Referring to the main topic, we clearly know that having three positive roots means that $a_{\rm{min}}+a_{\rm{max}}+a_{\rm{T}}>0$. It is common that this condition contradicts  the equation (\ref{Fbi30}). Therefore, we cannot obtain any potential with three positive roots for the matter dominated flat universe in the minimal massive bigravity model.\\

~~~~~~~~~~~~~~~~~~~~~~~~~~~~~~~~~~~~~~~~~~~~~~~~~~~~~~~~\textbf{2}. Two positive roots\\

In this case, we assume that $a_{1}<0$, and $0<a_{2}\leq a_{3}$, and set $a_{2}=a_{T_{1}}$ and $a_{3}=a_{T_{1}}$. Thus, comparing with (\ref{Fbi32}) it turns out that

\begin{align}\label{Fbi34}
a_{1}a_{T_{1}}a_{T_{2}}=\frac{\lambda}{3m^{2}M_{g}^{2}\left(\frac{3\beta_{2}^{2}}{\beta_{4}}-\frac{\beta_{0}}{3}\right)}<0.
\end{align}
Considering (\ref{Fbi32}) and the above inequality we obtain the result  $\lambda >0$. For sure, this condition corresponds to two deferent situations: two positive roots and one negative or three negative roots. In order to distinguish these two cases, we are supposed to consider the signs of $a_{1}+a_{\rm{min}}+a_{\rm{max}}$ and $a_{1}a_{\rm{min}}+a_{\rm{min}}a_{\rm{max}}+a_{1}a_{\rm{max}}$. Clearly, when there is at least one positive root, we have $a_{1}+a_{\rm{min}}+a_{\rm{max}}\geq 0$, which definitely corresponds to the case of two positive roots and one negative root. Here, we should notice that this result is in agreement with (\ref{Fbi30}). For $a_{1}<0<a_{\rm{min}}\leq a_{\rm{max}}$, we can reason that $a_{1}a_{\rm{min}}+a_{\rm{min}}a_{\rm{max}}+a_{1}a_{\rm{max}}=a_{1}\left(a_{\rm{min}}+a_{\rm{max}}\right)+a_{\rm{min}}a_{\rm{max}}<
-\left(a_{\rm{min}}+a_{\rm{max}}\right)^{2}+a_{\rm{min}}a_{\rm{max}}=-\left(a_{\rm{min}}+
\frac{1}{2}a_{\rm{max}}\right)^{2}-\frac{1}{4}a_{\rm{max}}^{2}<0$. However, for three negative roots we will have $a_{1}+a_{\rm{min}}+a_{\rm{max}}<0$ which contradicts  Eq.(\ref{Fbi30}).
So, for $0<\lambda<\lambda^{+}$ besides the inequality of (\ref{Fbi33}), we have shown the evolution of the potential $V(a)$ in Figure 2.

\begin{figure*}[ht]
  \centering
  \includegraphics[width=2.5in]{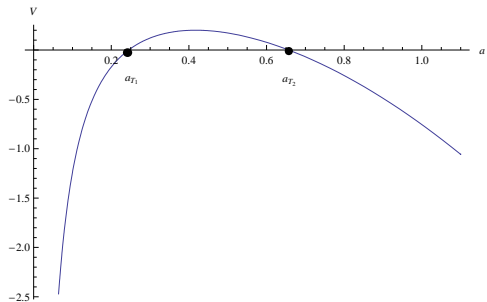}\hspace{2cm}
  \caption{The potential $V(a)$ for a BB $\Rightarrow$ BC universe or a bouncing one with model parameters satisfying $0<\lambda<\lambda^{+}$ and $\beta_{4}>\rm{Max} \{\frac{9\beta_{2}^{2}}{\beta_{0}},3\beta_{2}M_{*}^{2}\}$ and  the constants as $m=1$, $M_{g}=M_{f}$, $M_{*}=1$, $\beta_{0}=3$, $\beta_{2}=0.2$, $\beta_{4}=1.5$ and $\lambda=0.4$. The radii are $a_{T_{1}}=0.244685$ and $a_{T_{2}}=0.656934$.}
  \label{stable1}
\end{figure*}

Since the potential is negative within $a\in \left(0,a_{T_{1}}\right]$ and $\left[a_{T_{2}},\infty\right)$, and also $V(a)=0$  at $a=a_{T_{1}}$ and $a=a_{T_{2}}$, from the Figure 2, we conclude that a BB$\Rightarrow$BC universe or a bouncing one may rise. Therefore, if a big bang occurs at the initial moment, it may expand to $a_{T_{1}}$. It thereafter turns back at $a_{T_{1}}$ and terminates with a big crunch. Likewise, it is also possible that the universe experiences quantum tunneling from $a_{T_{1}}$ directly to $a_{T_{2}}$ and then expands forever. Additionally, if the universe contracts initially from infinity, the universe will have a bounce at $a_{T_{2}}$.\
While we reach the case $\lambda=\lambda^{+}$, $a_{T_{1}}$ and $a_{T_{2}}$ coincide with each other and give a double root $a_{\rm{s}}^{+}$ given in Equation (\ref{Fbi25}). The coefficients $\beta_{0}$, $\beta_{2}$ and $\beta_{4}$ must satisfy (\ref{Fbi33}), too. Figure 3. shows the Einstein static solution at $a_{\rm{s}}^{+}$ which is unstable. Thus, if we have a big bang singularity, it will expand to an unstable Einstein static universe and afterwards  turns over to a big crunch or expands forever. On the other hand, if the universe initially contracts from infinity, it can experience a bounce at $a_{\rm{s}}^{+}$ or end up with a big crunch.\\\\

\begin{figure*}[ht]
  \centering
  \includegraphics[width=2.5in]{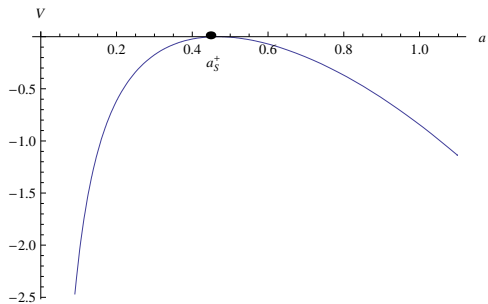}\hspace{2cm}
  \caption{The potential $V(a)$ with model parameters satisfying $\lambda=\lambda^{+}$ and $\beta_{4}>\rm{Max} \{\frac{9\beta_{2}^{2}}{\beta_{0}},3\beta_{2}M_{*}^{2}\}$. An unstable Einstein static solution and a big bang universe are obtained
using the constants as $m=1$, $M_{g}=M_{f}$, $M_{*}=1$, $\beta_{0}=3$, $\beta_{2}=0.2$, $\beta_{4}=1.5$ and $\lambda=0.5595$ with the radii $a_{\rm{s}}^{+}=0.4600$.}
  \label{stable2}
\end{figure*}

~~~~~~~~~~~~~~~~~~~~~~~~~~~~~~~~~~~~~~~~~~~~~~~~~~~~~~~~~~\textbf{3}. One positive roots\\

In this case, we assume that $a_{1}$, $a_{2}<0$, and $a_{3}=a_{\rm{T}}>0$, therefore we require
\begin{align}\label{Fbi35}
a_{1}a_{2}a_{\rm{T}}>0,
\end{align}
from which, together with the relation (\ref{Fbi33}) and  $\left(\frac{-3\beta_{2}^{2}}{\beta_{4}}+\frac{\beta_{0}}{3}\right)>0$ one finds
\begin{align}\label{Fbi36}
\lambda<0.
\end{align}
So, the final allowed interval for $\lambda$ becomes
\begin{align}\label{Fbi37}
\lambda^{-}\leq\lambda<0.
\end{align}

Clearly, this result is not acceptable since we should deal with positive $\lambda$s.\\

~~~~~~~~~~~~~~~~~~~~~~~~~~~~~~~~~~~~~~~~~~~~~~~~~~~~~~~~~~\textbf{4}. No positive roots\\

 According to the condition (\ref{Fbi30}), $V(a)=0$ is not allowed to have three negative roots because the summation $a_{1}+a_{2}+a_{\rm{T}}$ should be always negative.\\

~~~~~~~~~~~~~~~~~~~~~~~~~~~~~~~~~~~~~~~~~~~~~~~~~~~~~~~~~~~~~~~~\textbf{B}. $\lambda>\lambda^{+}$\\

Exerting this condition, there is only one real root (which can be positive or negative) with other two roots as conjugate imaginary pair. Now, we are going to consider again the following classifications. \\

~~~~~~~~~~~~~~~~~~~~~~~~~~~~~~~~~~~~~~~~~~~~~~~~~~~~~~~~~~\textbf{1}. One positive root\\

Under the assumption that $a_{1}$ and $a_{2}$ are two conjugate imaginary roots ($a_{2}=a_{1}^{*}$) and $a_{3}$ is the only positive one, the condition (\ref{Fbi32}) reads

\begin{align}\label{Fbi38}
a_{1}a_{2}a_{3}=\frac{\lambda}{3m^{2}M_{g}^{2}\left(\frac{3\beta_{2}^{2}}{\beta_{4}}-\frac{\beta_{0}}{3}\right)}>0.
\end{align}
Since $\left(\frac{3\beta_{2}^{2}}{\beta_{4}}-\frac{\beta_{0}}{3}\right)<0$, we have

\begin{align}\label{Fbi39}
\lambda<0,
\end{align}
which is clearly in contradiction with $\lambda>\lambda^{+}$. As a result, this case is an impossible one.\\

~~~~~~~~~~~~~~~~~~~~~~~~~~~~~~~~~~~~~~~~~~~~~~~~~~~~~~~~~~~\textbf{2}. No positive root\\

This case contains just one negative real root which gives

\begin{align}\label{Fbi40}
a_{1}a_{2}a_{3}=\frac{\lambda}{3m^{2}M_{g}^{2}\left(\frac{3\beta_{2}^{2}}{\beta_{4}}-\frac{\beta_{0}}{3}\right)}<0,
\end{align}

from which we can write

\begin{align}\label{Fbi41}
\lambda>0.
\end{align}

It leads to the following final result

\begin{align}\label{Fbi42}
\lambda>\lambda^{+}.
\end{align}

From the above inequality and the previous condition given in equation (\ref{Fbi33}), we can plot Figure. 4 in which the potential is always negative and the type of cosmic evolution is BB$\Rightarrow \infty$ or $\infty \Rightarrow$ BC.\\

\begin{figure*}[ht]
  \centering
  \includegraphics[width=2.5in]{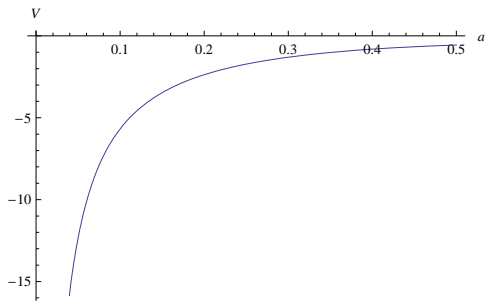}\hspace{2cm}
  \caption{The potential $V(a)$ for a BB $\Rightarrow \infty$ or $\infty \Rightarrow$ BC universe  with model parameters satisfying $\lambda>\lambda^{+}$ and $\beta_{4}>\rm{Max} \{\frac{9\beta_{2}^{2}}{\beta_{0}},3\beta_{2}M_{*}^{2}\}$
with the constants as $m=1$, $M_{g}=M_{f}$, $M_{*}=1$, $\beta_{0}=2.1$, $\beta_{2}=0.5$, $\beta_{4}=3$ and $\lambda=1$.}
  \label{stable3}
\end{figure*}

Before going to the open universe case with $\kappa=-1$, a remark is in order related to the coefficient of $a^{3}$ term, namely $\frac{\frac{-3\beta_{2}^{2}}{\beta_{4}}+\frac{\beta_{0}}{3}}{\frac{3M_{*}^{2}\beta_{2}}{\beta_{4}}-1}$, in the potential (\ref{Fbi29}). The equality $\frac{\beta_{0}}{3}=\frac{3\beta_{2}^{2}}{\beta_{4}}$ in the nominator will never happen because it gives a linear potential with constant derivative.\\

\textbf{Case II:} $\kappa=-1$\\

We have the potential as

\be\label{Fbi43}
V(a)=\frac{a^2 m^2 \left(-\frac{3 \beta _2^2}{\beta _4}+\frac{\beta _0}{3}+\frac{\lambda }{3 a^3 m^2 M_g^2}\right)}{\frac{3 \beta _2 M_*^2}{\beta _4}-1}-1,
\ee
and then
\be\label{Fbi44}
V'(a)=\frac{2 a m^2 \left(-\frac{3 \beta _2^2}{\beta _4}+\frac{\beta _0}{3}+\frac{\lambda }{3 a^3 m^2 M_g^2}\right)}{\frac{3 \beta _2 M_*^2}{\beta _4}-1}-\frac{\lambda
   }{a^2 M_g^2 \left(\frac{3 \beta _2 M_*^2}{\beta _4}-1\right)}.
\ee
We have the following expressions for $\lambda^{-}$, $a_{\rm{S}}^{+}$ and $a_{\rm{T}}^{\mp}$
\be\label{Fbi45}
\lambda=\lambda^{-}= \frac{-2M_{g}^{2}\left(-\beta_{4}+3\beta_{2}M_{*}^{2}\right)^{\frac{3}{2}}}{\beta_{4}\sqrt{m^{2}
\left(-9\beta_{2}^{2}+\beta_{0}\beta_{4}\right)}},
\ee

\be\label{Fbi46}
a_{\rm{s}}=a_{\rm{s}}^{+}= \frac{\sqrt{-\beta_{4}+3\beta_{2}M_{*}^{2}}}{\sqrt{m^{2}
\left(-9\beta_{2}^{2}+\beta_{0}\beta_{4}\right)}},
\ee
and
\be\label{Fbi47}
a_{\rm{T}}=a_{\rm{T}}^{\mp}=-\frac{m^{4}\left(-9\beta_{2}^{2}+\beta_{0}\beta_{4}\right)^{2}\left(-\beta_{4}+3\beta_{2}M_{*}^{2}\right)
+\left( \mp m^{6}\left(9\beta_{2}^{2}-\beta_{0}\beta_{4}\right)^{3}\left(-\beta_{4}+3\beta_{2}M_{*}^{2}\right)^{\frac{3}{2}}\right)^{\frac{2}{3}}}{
m^{3}\left(-9\beta_{2}^{2}+\beta_{0}\beta_{4}\right)^{\frac{3}{2}}\left( \mp m^{6}\left(9\beta_{2}^{2}-\beta_{0}\beta_{4}\right)^{3}\left(-\beta_{4}+3\beta_{2}M_{*}^{2}\right)^{\frac{3}{2}}\right)^{\frac{1}{3}}}.
\ee
Since the terms $\lambda^{-}$, $a_{\rm{S}}^{+}$ and $a_{\rm{T}}^{\mp}$ should
be real, we have to limit ourselves to the following ranges

\begin{align}\label{Fbi48}
3\beta_{2}M_{*}^{2}<\beta_{4}<\frac{9\beta_{2}^{2}}{\beta_{0}},
\end{align}
which for the fixed $\beta_{0}=2$ is plotted in Fig 5.

 \begin{figure*}[ht]
  \centering
  \includegraphics[width=2.5in]{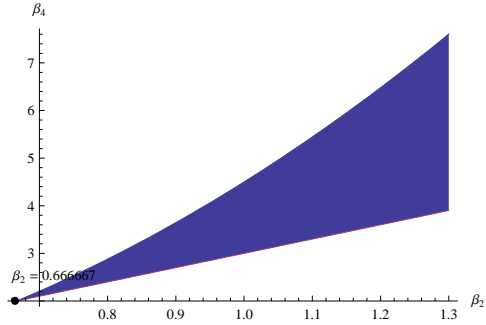}\hspace{2cm}
  \caption{Phase diagram of space-times in $\left(\beta_{2},\beta_{4}\right)$ plane when $\gamma$, $\lambda^{-}$, $a_{\rm{S}}^{+}$ and $a_{\rm{T}}^{-}$ are real in colored region.}
  \label{stable4}
\end{figure*}

Similar to the relation (\ref{Fbi27}), the potential (\ref{Fbi43})can be rewritten as follows
\be\label{Fbi49}
V(a)=\frac{m^{2}\left(\frac{-3\beta_{2}^{2}}{\beta_{4}}+\frac{\beta_{0}}{3}\right)}{a\left(\frac{3M_{*}^{2}\beta_{2}}{\beta_{4}}-1\right)}\times
\left(a^{3}+\frac{\left(-\frac{3M_{*}^{2}\beta_{2}}{\beta_{4}}+1\right)a}{m^{2}\left(\frac{-3\beta_{2}^{2}}{\beta_{4}}+\frac{\beta_{0}}{3}\right)}
+\frac{\lambda}{3m^{2}M_{g}^{2}\left(\frac{-3\beta_{2}^{2}}{\beta_{4}}+\frac{\beta_{0}}{3}\right)}\right).
\ee
In this sense, we would like to notice that the coefficient of $a^{3}$ term is the same as that of the case I. Thus, under the conditions mentioned in (\ref{Fbi49}) the term $\frac{\frac{-3\beta_{2}^{2}}{\beta_{4}}+\frac{\beta_{0}}{3}}{\frac{3M_{*}^{2}\beta_{2}}{\beta_{4}}-1}$ is positive-valued.
As in case I, here we also have the following classifications A and B as\\

~~~~~~~~~~~~~~~~~~~~~~~~~~~~~~~~~~~~~~~~~~~~~~~~~~~~~~~~~~~\textbf{A}. $0\leq \lambda \leq \lambda^{-}$.\\

We see that imposing $V(a)=0$ constructs a cubic equation of $a$ with three real roots $a_{1}$, $a_{2}$ and $a_{3}$. As a result, the equation (\ref{Fbi49}) is rewritten as
\begin{align}\label{Fbi50}
V(a)=&
\frac{m^{2}\left(\frac{-3\beta_{2}^{2}}{\beta_{4}}+\frac{\beta_{0}}{3}\right)}{a\left(\frac{3M_{*}^{2}\beta_{2}}{\beta_{4}}-1\right)}
\left(a-a_{\rm{min}}\right)\left(a-a_{\rm{max}}\right)\left(a-a_{\rm{T}}\right)=
\frac{m^{2}\left(\frac{-3\beta_{2}^{2}}{\beta_{4}}+\frac{\beta_{0}}{3}\right)}{a\left(\frac{3M_{*}^{2}\beta_{2}}{\beta_{4}}-1\right)}\times \nn&
\left(a^{3}-\left(a_{\rm{min}}+a_{\rm{max}}+a_{\rm{T}}\right)a^{2}+\left(a_{\rm{min}}a_{\rm{max}}+a_{\rm{T}}a_{\rm{min}}+
a_{\rm{max}}a_{\rm{T}}\right)a-a_{\rm{min}}a_{\rm{max}}a_{\rm{T}}\right).
\end{align}
Now, the comparison of (\ref{Fbi50}) with (\ref{Fbi49}) gives us
\begin{align}\label{Fbi51}
a_{\rm{min}}+a_{\rm{max}}+a_{\rm{T}}=0,
\end{align}

\begin{align}\label{Fbi52}
a_{\rm{min}}a_{\rm{max}}+a_{\rm{T}}a_{\rm{min}}+a_{\rm{max}}a_{\rm{T}}=\frac{\left(\frac{-3M_{*}^{2}\beta_{2}}{\beta_{4}}+1\right)}{
m^{2}\left(\frac{-3\beta_{2}^{2}}{\beta_{4}}+\frac{\beta_{0}}{3}\right)},
\end{align}
and

\begin{align}\label{Fbi53}
a_{\rm{min}}a_{\rm{max}}a_{\rm{T}}=\frac{\lambda}{3m^{2}M_{g}^{2}\left(\frac{3\beta_{2}^{2}}{\beta_{4}}-\frac{\beta_{0}}{3}\right)},
\end{align}
where $a_{1}=a_{\rm{min}}$, $a_{2}=a_{\rm{max}}$ and $a_{3}=a_{\rm{T}}$.
As  mentioned before, three positive roots case will not be possible since the summation of three roots in (\ref{Fbi51}) should be zero. Therefore, we consider two positive roots case.\\

~~~~~~~~~~~~~~~~~~~~~~~~~~~~~~~~~~~~~~~~~~~~~~~~~~~~~~~\textbf{1}. Two positive roots\\

Under the assumption $a_{1}<0<a_{\rm{min}} \leq a_{\rm{max}}$, we require that

\begin{align}\label{Fbi54}
a_{1}a_{\rm{min}}a_{\rm{max}}=\frac{\lambda}{3m^{2}M_{g}^{2}\left(\frac{3\beta_{2}^{2}}{\beta_{4}}-\frac{\beta_{0}}{3}\right)}<0,
\end{align}

from which, besides the condition (\ref{Fbi48}), we imply that $\lambda$ should be negative. Thus we leave this part here because we should work with negative $\lambda$.

~~~~~~~~~~~~~~~~~~~~~~~~~~~~~~~~~~~~~~~~~~~~~~~~~~~~~~~\textbf{2}. One positive root\\

Under the assumption $a_{1},a_{2}<0$ and $a_{3}=a_{\rm{T}}>0$, we conclude that
\begin{align}\label{Fbi55}
a_{1}a_{2}a_{\rm{T}}=\frac{\lambda}{3m^{2}M_{g}^{2}\left(\frac{3\beta_{2}^{2}}{\beta_{4}}-\frac{\beta_{0}}{3}\right)}>0,
\end{align}
from which, besides the conditions (\ref{Fbi57}), one can imply that
\begin{align}\label{Fbi56}
0<\lambda \leq \lambda^{-}.
\end{align}
By obeying the conditions $0<\lambda\leq\lambda^{-}$ and (\ref{Fbi48}), we have plotted the evolution of $V(a)$ in Figure 6. We have found that a big bang to big crunch evolution is appeared because $V(a)\leq 0$ in $(0,a_{\rm{T}}]$ with $V(a)=0$ happening at $a=a_{\rm{T}}$.

\begin{figure*}[ht]
  \centering
  \includegraphics[width=2.5in]{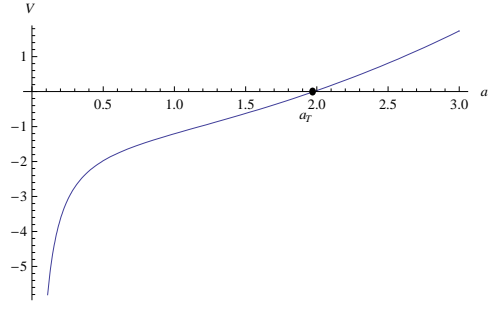}\hspace{2cm}
  \caption{The potential $V(a)$ for a BB $\Longrightarrow$ BC universe  with model parameters satisfying $0<\lambda\leq\lambda^{-}$ and $3\beta_{2}M_{*}^{2}<\beta_{4}<\frac{9\beta_{2}^{2}}{\beta_{0}}$
with the constants as $m=1$, $M_{g}=M_{f}$, $M_{*}=1$, $\beta_{0}=2$, $\beta_{2}=1$, $\beta_{4}=4.01$, $\lambda=0.4$, $a_{\rm{T}}=1.97956$.}
  \label{stable5}
\end{figure*}

Eventually, before we finish this subsection we note that the case ``no positive root" is rejected because it yields $a_{1}+a_{2}+a_{3}<0$ which violates (\ref{Fbi51}).\\

~~~~~~~~~~~~~~~~~~~~~~~~~~~~~~~~~~~~~~~~~~~~~~~~~~~~~~~~~~~\textbf{B}. $\lambda>\lambda^{-}$\\

As explained before, in previous subsections, under this condition we only have one real root $a_{3}$ and the other two roots are a conjugate imaginary pair. Now, we are allowed to have  two following classifications.\\\\\\

~~~~~~~~~~~~~~~~~~~~~~~~~~~~~~~~~~~~~~~~~~~~~~~~~~~~~~~\textbf{1}. One positive root\\

Obviously, the product of these three roots will be positive. Therefore, according to the equation (\ref{Fbi55}) and (\ref{Fbi48}) we obtain
\begin{align}\label{Fbi57}
\lambda>\lambda^{-}.
\end{align}
Therefore, by using the above inequality and the allowed ranges in (\ref{Fbi48}), we have plotted Figure 7. in which we face with a \textcolor[rgb]{1,0,0.501961}{BB $\Longrightarrow$ BC}  universe.

\begin{figure*}[ht]
  \centering
  \includegraphics[width=2.5in]{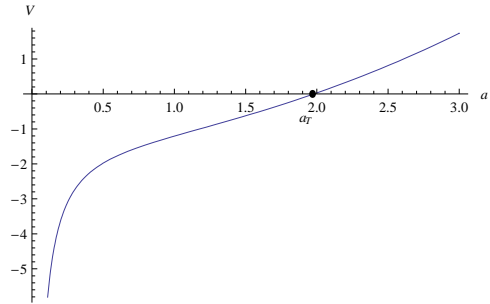}\hspace{2cm}
  \caption{The potential $V(a)$ for a BB $\Longrightarrow$ BC universe  with model parameters satisfying $\lambda>\lambda^{-}$ and $3\beta_{2}M_{*}^{2}<\beta_{4}<\frac{9\beta_{2}^{2}}{\beta_{0}}$ with the constants as $m=1$, $M_{g}=M_{f}$, $M_{*}=1$, $\beta_{0}=2$, $\beta_{2}=1$, $\beta_{4}=4.01$, $\lambda=0.6$ and $a_{\rm{T}}=2.906851$.}
  \label{stable6}
\end{figure*}

Here, we should note that the part ``no positive root" will not be mentioned because it results in $\lambda<0$ which is not acceptable.\\

\textbf{Case III:} $\kappa=0$\\

  Essentially, considering the current observations
on the universe, we are mostly motivated to  study the $\kappa=0$ case,  but in the $\beta_{1}=\beta_{3}=0$ massive bigravity model with the potential (\ref{Fbi29}) for the matter-dominated and radiation-dominated cases, we cannot find any analytical solutions for $a$ and $\lambda$ in terms of the free parameters $\beta_{i}$'s. To more clarification, we refer to the potential (\ref{Fbi21})  with  $\omega=0$ and $\kappa=0$ for the matter-dominated flat space universe as follows

 \be\label{Fbi58}
V(a)=m^{2}a^{2}\times\frac{\frac{-3\beta_{2}^{2}}{\beta_{4}}+\frac{\beta_{0}}{3}+\frac{\lambda}{3a^{3}m^{2}M_{g}^{2}}}{
\frac{3M_{*}^{2}\beta_{2}}{\beta_{4}}-1}.
\ee

Under the condition of ES state we can write

 \be\label{Fbi59}
V(a)=0,
\ee

from which we obtain $\lambda$ as

 \be\label{Fbi60}
\lambda=\frac{a^{3}M_{g}^{2}m^{2}\left(9\beta_{2}^{2}-\beta_{0}\beta_{4}\right)}{\beta_{4}}.
\ee

Another ES state condition is $V'(a)=0$ which reads as

 \be\label{Fbi61}
\frac{2 m^2 M_g^2 \left(9 \beta _2^2-\beta _0 \beta _4\right) a^3+\lambda  \beta _4}{3 a^2 M_g^2 \left(\beta _4-3 \beta _2 M_*^2\right)}=0.
\ee

Putting (\ref{Fbi60}) into the above equation, we see that all terms cancel each other and we are not  able to find any relation for $a$ or $\lambda$ in terms of the free parameters $\beta_{i}$'s. Thus, we have to study this case numerically by means of plotting the Figure 8.\\

\begin{figure*}[ht]
  \centering
  \includegraphics[width=2.5in]{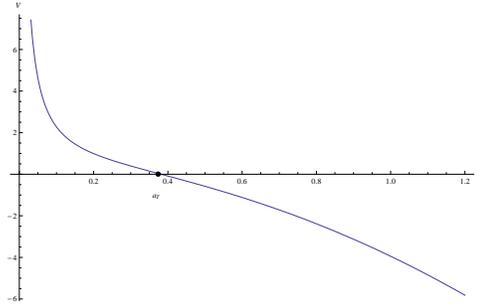}\hspace{2cm}
  \caption{The potential $V(a)$ becomes negative for $a\in [a_{T},\infty)$  describing the bounce one with the constants as $m=1$, $M_{g}=M_{f}$, $M_{*}=1$, $\beta_{0}=1$, $\beta_{2}=2.6$, $\beta_{4}=3.2$ and $\lambda=1$. The  bouncing radii is $a_{T}=0.375$.}
  \label{stable7}
\end{figure*}

\section{THE EVOLUTION OF A RADIATION-DOMINATED UNIVERSE IN THE MASSIVE BIGRAVITY MODEL}

In this section, we consider the case where the universe is dominated by radiation with $\omega=\frac{1}{3}$ for close and open universes. Thus, the cosmic energy density can be proposed by $\rho=\frac{\lambda}{a^{4}}$.\\

\textbf{Case I}: $\kappa=1$ \\

The potential resulted from the assumption $\dot{a}=0$ which belongs to the ES solution becomes
\begin{align}\label{Fbi62}
V(a)=-\frac{\lambda}{3M_{g}^{2}a^{2}}-\frac{1}{3}a^{2}m^{2}\beta_{0}+\frac{3a^{2}m^{2}\beta_{2}^{2}+\beta_{4}-3\beta_{2}M_{*}^{2}}{\beta_{4}}.
\end{align}
Similar to the previous section, combining $V(a)=0$ and $V'(a)=0$ we find $\lambda$ and the Einstein static solutions as follows
\be\label{Fbi63}
\lambda=\frac{9M_{g}^{2}\left(\beta_{4}-3\beta_{2}M_{*}^{2}\right)^{2}}{4m^{2}\beta_{4}\left(-9\beta_{2}^{2}+\beta_{0}\beta_{4}\right)},
\ee
\be\label{Fbi64}
a_{\rm{s}}^{\pm}=a_{\rm{T}}^{\pm}=\pm \frac{\sqrt{\frac{3}{2}\left(\beta_{4}-3\beta_{2}M_{*}^{2}\right)}}{\sqrt{m^{2}\left(-9\beta_{2}^{2}+\beta_{0}\beta_{4}\right)}}.
\ee
Obviously, the radiation-dominated potential has four roots. Meantime, we should extract ranges that the solutions are real-valued, so we find the following inequality

\begin{align}\label{Fbi65}
\beta_{4}>\rm{Max} \left\{\frac{9\beta_{2}^{2}}{\beta_{0}},3\beta_{2}M_{*}^{2}\right\}.
\end{align}

Moreover, by using the equation (\ref{Fbi21}), the general form of the potential for radiation-dominated universe is obtained as
\be\label{Fbi66}
V(a)=1+m^{2}a^{2}\times\frac{\frac{-3\beta_{2}^{2}}{\beta_{4}}+\frac{\beta_{0}}{3}+\frac{\lambda}{3a^{4}m^{2}M_{g}^{2}}}{
\frac{3M_{*}^{2}\beta_{2}}{\beta_{4}}-1}.
\ee
This can be rewritten in the following form
\be\label{Fbi67}
V(a)=\frac{m^{2}\left(\frac{-3\beta_{2}^{2}}{\beta_{4}}+\frac{\beta_{0}}{3}\right)}{a^{2}\left(\frac{3M_{*}^{2}\beta_{2}}{\beta_{4}}-1\right)}\times
\left(a^{4}+\frac{\left(\frac{3M_{*}^{2}\beta_{2}}{\beta_{4}}-1\right)a^{2}}{m^{2}\left(\frac{-3\beta_{2}^{2}}{\beta_{4}}+\frac{\beta_{0}}{3}\right)}
+\frac{\lambda}{3m^{2}M_{g}^{2}\left(\frac{-3\beta_{2}^{2}}{\beta_{4}}+\frac{\beta_{0}}{3}\right)}\right),
\ee
which clearly shows the forth order of the scale factor in the potential. As a result, we have found an unstable static universe in which the universe is born from a big bang singularity and then expands to an Einstein static universe ($V(a_{\rm{S}})=0$) which is unstable and may bounce to a big crunch. The potential behavior is shown in Figure 9.\\
\begin{figure*}[ht]
  \centering
  \includegraphics[width=2.5in]{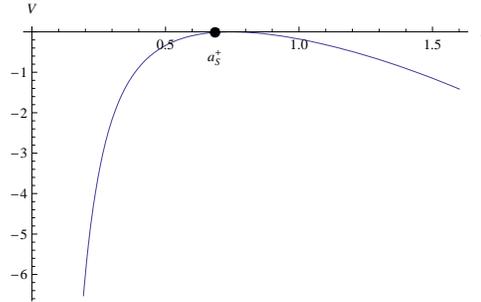}\hspace{2cm}
  \caption{The potential $V(a)$  with model parameters satisfying  $\beta_{4}>\rm{Max} \{\frac{9\beta_{2}^{2}}{\beta_{0}},3\beta_{2}M_{*}^{2}\} $ turns out an unstable Einstein static universe with the constants as $m=1$, $M_{g}=M_{f}$, $M_{*}=1$, $\beta_{0}=2.1$, $\beta_{2}=0.5$, $\beta_{4}=3$, $\lambda=0.416667$ and with the radii  $a_{\rm{S}}^{+}=a_{\rm{T}}^{+}=0.7$.}
  \label{stable8}
\end{figure*}

\textbf{Case II}: $\kappa=-1$ \\

In an open radiation-dominated universe, the Einstein static potential is obtained as

\begin{align}\label{Fbi68}
V(a)=-\frac{\lambda}{3M_{g}^{2}a^{2}}-\frac{1}{3}a^{2}m^{2}\beta_{0}+\frac{3a^{2}m^{2}\beta_{2}^{2}-\beta_{4}+3\beta_{2}M_{*}^{2}}{\beta_{4}}.
\end{align}
Again, upon the conditions $V(a)=V'(a)=0$, we find
\be\label{Fbi69}
\lambda=\frac{9M_{g}^{2}\left(\beta _4+3  \beta _2 M_*^2\right){}^2}{4 m^2 \beta _4 \left(\beta _0 \beta _4-9 \beta _2^2\right)},
\ee
\be\label{Fbi70}
a^{\pm}=\pm \frac{\sqrt{\frac{3}{2}} \sqrt{3 M_*^2 \beta _2-\beta _4}}{\sqrt{m^2 \left(\beta _0 \beta _4-9 \beta _2^2\right)}}.
\ee

Clearly, the requirement to have a real-valued $a$ leads to  a negative $\lambda$. Thus, we can say that we do not have any radiation-dominated open universe evolution in the early time in the minimal massive bigravity model.\\

\textbf{Case III}:  $\kappa=0$ \\

Similar to what we have explained in the matter-dominated spatially flat  universe (see equations (\ref{Fbi57}) to (\ref{Fbi61})),  we consider the potential (\ref{Fbi21}) with $\omega=\frac{1}{3}$ and $\kappa=0$  as follows

\begin{align}\label{Fbi71}
V(a)=\frac{a^2 m^2 \left(-\frac{3 \beta _2^2}{\beta _4}+\frac{\beta _0}{3}+\frac{\lambda }{3 a^4 m^2 M_g^2}\right)}{\frac{3 \beta _2 M_*^2}{\beta _4}-1}.
\end{align}
For $V(a)=0$, we can find

\begin{align}\label{Fbi72}
\lambda=\frac{a^4 M_g^2 m^{2}\left(9  \beta _2^2-\beta _0 \beta _4\right)}{\beta _4}.
\end{align}
Inserting $\lambda$ in $V'(a)=0$, we obtain $0=0$ which dose not help us to obtain any expression for $\lambda$ and $a$ in terms of $\beta_{i}$'s. Therefore, we proceed with the following numerical analysis via plotting Figure 10.

\begin{figure*}[ht]
  \centering
  \includegraphics[width=2.5in]{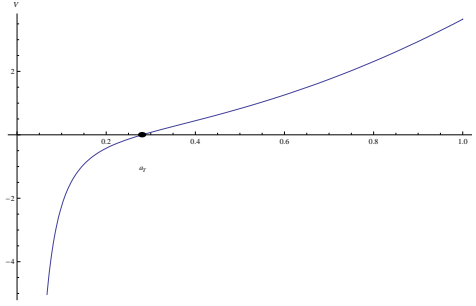}\hspace{2cm}
  \caption{The potential $V(a)$ becomes negative for $a\in (\infty,a_{T}]$ describing the BB $\Rightarrow$ BC evolution with the constants as $m=1$, $M_{g}=M_{f}$, $M_{*}=1$, $\beta_{0}=1$, $\beta_{2}=-4.1$, $\beta_{4}=1.33$ and $\lambda=0.7$. The  bouncing radii is $a_{T}=0.28$.}
  \label{stable9}
\end{figure*}

\section{THE EVOLUTION OF A UNIVERSE With DUST AND PHANTOM IN THE MASSIVE BIGRAVITY MODEL}

According to what we have done in two previous sections, we could not find an oscillating universe. In order to find such solutions, we may examine a combination of cosmic energy densities of dust and phantom with $\omega=0$ and $\omega=-\frac{2}{3}$, respectively. Actually, this choice makes it possible to have a potential $V(a)$ with three positive roots evolving in a shape that gives us the opportunity to obtain an oscillating universe and a stable ES state. We explain this idea in details in this remaining section. Thus, we consider an open and also close universes dominated by dust and phantom.\\

\textbf{Case I}: $\kappa=1$ \\

By this assumption, the potential (\ref{Fbi21}) takes the following form
\be\label{Fbi73}
V(a)=1+m^{2}a^{2}\times\frac{\frac{-3\beta_{2}^{2}}{\beta_{4}}+\frac{\beta_{0}}{3}+\frac{\lambda}{3m^{2}M_{g}^{2}}\left(\frac{1}{a^{3}}+\frac{1}{a}\right)}{
\frac{3M_{*}^{2}\beta_{2}}{\beta_{4}}-1}.
\ee

Under the conditions $V(a)=0$ and $V'(a)=0$, we obtain the following Einstein static solutions
\begin{eqnarray}\label{eq74}
\lambda^{\pm}&=&\frac{1}{4\sqrt{2}\beta_{4}\sqrt{m^{2}\left(9\beta_{2}^{2}-\beta_{0}\beta_{4}\right)}}M_{g}^{2}\left(27 m^2 \beta _2^2-9 M_*^2 \beta _2+\left(3-3 m^2 \beta _0\right) \beta _4\right.\nonumber\\&&\left.\pm\sqrt{3} \sqrt{\left(27 m^2 \beta _2^2+3 M_*^2 \beta _2-\left(3 \beta _0 m^2+1\right)
   \beta _4\right) \left(9 m^2 \beta _2^2+9 M_*^2 \beta _2-\left(\beta _0 m^2+3\right) \beta _4\right)}\right)\times \nonumber\\&& \left(-27 m^2 \beta _2^2-9 M_*^2 \beta _2+3 \left(\beta _0 m^2+1\right) \beta _4\right.\nonumber\\&&\left.\mp\sqrt{3} \sqrt{\left(27 m^2 \beta _2^2+3 M_*^2 \beta _2-\left(3 \beta _0 m^2+1\right)
   \beta _4\right) \left(9 m^2 \beta _2^2+9 M_*^2 \beta _2-\left(\beta _0 m^2+3\right) \beta _4\right)}\right)^{\frac{1}{2}}
\end{eqnarray}

\be\label{Fbi75}
a_{\rm{S}}^{\pm}=\frac{\sqrt{\frac{-27 m^2 \beta _2^2-9 M_*^2 \beta _2+3 \left(\beta _0 m^2+1\right) \beta _4\pm\sqrt{3} \sqrt{\left(27 m^2 \beta _2^2+3 M_*^2 \beta _2-\left(3 \beta _0
   m^2+1\right) \beta _4\right) \left(9 m^2 \beta _2^2+9 M_*^2 \beta _2-\left(\beta _0 m^2+3\right) \beta _4\right)}}{m^2 \left(9 \beta _2^2-\beta _0 \beta
   _4\right)}}}{\sqrt{2}}.
\ee

Deriving the explicit form of $a_{\rm{T}}^{\pm}$, we obtain a very large output that makes it impossible to be written here. Consequently, we are supposed to consider the allowed ranges in which the above terms become real and positive. Therefore, we reach the following result

\be\label{Fbi76}
\rm{Max}\left\{3\beta_{2}M_*^2 ,\frac{3\beta_{2}M_*^2\left(1+9\beta_{2}M_*^2\right)}{1+3\beta_{0}}\right\}<\beta_{4}<\frac{9\beta_{2}^{2}}{\beta_{0}}.
\ee

Fixing $\beta_{0}=1$ (because it actually plays the role of a cosmological constant of the metric $g_{\mu\nu}$ and it is allowed to take an arbitrary positive value) we are left with the following condition
\be\label{Fbi77}
\frac{3\beta_{2}M_*^2}{4}\left(1+9\beta_{2}M_*^2\right)<\beta_{4}<\frac{9\beta_{2}^{2}}{\beta_{0}},
\ee
by which we have plotted the allowed ranges for the free parameters $\beta_{2}$ and $\beta_{4}$ in Figure 11.
\begin{figure*}[ht]
  \centering
  \includegraphics[width=2.5in]{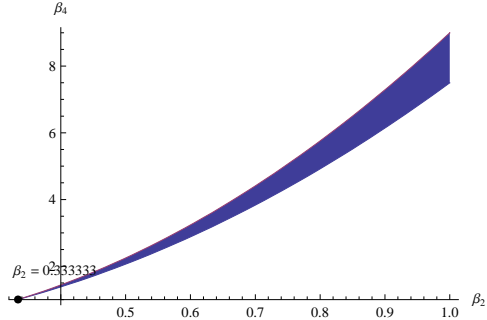}\hspace{2cm}
  \caption{Phase diagram of space-times in $\left(\beta_{2},\beta_{4}\right)$ plane when $\gamma$, $\lambda^{\pm}$, $a_{\rm{S}}^{\pm}$ and $a_{\rm{T}}^{\pm}$ are positive and real in colored region.}
  \label{stable10}
\end{figure*}
Rewriting the potential (\ref{Fbi73}) in the following form
\be\label{Fbi78}
V(a)=\frac{m^{2}\left(\frac{-3\beta_{2}^{2}}{\beta_{4}}+\frac{\beta_{0}}{3}\right)}{a\left(\frac{3M_{*}^{2}\beta_{2}}{\beta_{4}}-1\right)}\times
\left(a^{3}+\frac{\lambda a^{2}}{3m^{2}M_{g}^{2}\left(\frac{-3\beta_{2}^{2}}{\beta_{4}}+\frac{\beta_{0}}{3}\right)}+
\frac{\left(\frac{3M_{*}^{2}\beta_{2}}{\beta_{4}}-1\right)a}{m^{2}\left(\frac{-3\beta_{2}^{2}}{\beta_{4}}+\frac{\beta_{0}}{3}\right)}
+\frac{\lambda}{3m^{2}M_{g}^{2}\left(\frac{-3\beta_{2}^{2}}{\beta_{4}}+\frac{\beta_{0}}{3}\right)}\right),
\ee
and comparing it with the relation (\ref{Fbi58}), we conclude that

\begin{align}\label{Fbi79}
a_{\rm{min}}+a_{\rm{max}}+a_{\rm{T}}=\frac{-\lambda}{3m^{2}M_{g}^{2}\left(\frac{-3\beta_{2}^{2}}{\beta_{4}}+\frac{\beta_{0}}{3}\right)},
\end{align}

\begin{align}\label{Fbi80}
a_{\rm{min}}a_{\rm{max}}+a_{\rm{T}}a_{\rm{min}}+a_{\rm{max}}a_{\rm{T}}=\frac{\left(\frac{3M_{*}^{2}\beta_{2}}{\beta_{4}}-1\right)}{
m^{2}\left(\frac{-3\beta_{2}^{2}}{\beta_{4}}+\frac{\beta_{0}}{3}\right)},
\end{align}
and

\begin{align}\label{Fbi81}
a_{\rm{min}}a_{\rm{max}}a_{\rm{T}}=\frac{-\lambda}{3m^{2}M_{g}^{2}\left(\frac{-3\beta_{2}^{2}}{\beta_{4}}+\frac{\beta_{0}}{3}\right)}.
\end{align}
We may have the following classifications\\

~~~~~~~~~~~~~~~~~~~~~~~~~~~~~~~~~~~~~~~~~~~~~~~~~~~~~~~~~~~~~~~\textbf{A}. $\lambda^{-}\leq \lambda \leq \lambda^{+}$\\

In spite of the previous sections, the matter-phantom-dominated universe provides a situation that the potential can have three positive roots since the summation of these roots (\ref{Fbi79}) becomes non-zero. In fact, when we say we have three positive roots, we mean that $a_{\rm{min}}+a_{\rm{max}}+a_{\rm{T}}>0$, $a_{\rm{min}}a_{\rm{max}}+a_{\rm{T}}a_{\rm{min}}+a_{\rm{max}}a_{\rm{T}}>0$ and $a_{\rm{min}}a_{\rm{max}}a_{\rm{T}}>0$ which confirms completely the inequality (\ref{Fbi76}). Thus, we should have $\lambda>0$ which means that the $\lambda^{+}$ and $\lambda^{-}$ should be positive. Therefore, we begin with the classification ``Three positive roots".\\

~~~~~~~~~~~~~~~~~~~~~~~~~~~~~~~~~~~~~~~~~~~~~~~~~~~~~~~~~\textbf{2}. Three positive roots\\

Assuming $a_{1}=a_{\rm{T}}$, $a_{2}=a_{\rm{min}}$ and $a_{3}=a_{\rm{max}}$ and also letting $0\leq a_{\rm{T}}\leq a_{\rm{min}}\leq a_{\rm{max}}$,
and according to the colored region of Figure 9. for $\lambda^{-}< \lambda < \lambda^{+}$, we plot the Figure 12. in which $V(a)\leq 0$ in $a\in [a_{\rm{min}},a_{\rm{max}}]$ while the equality happening at $a=a_{\rm{min}}$ and $a=a_{\rm{max}}$, and $a\in (0,a_{\rm{T}}]$ with $V(a_{\rm{T}})=0$, which leads to the universe oscillations between $a_{\rm{min}}$ and $a_{\rm{max}}$ or bouncing at $a_{\rm{T}}$. The cosmic evolution type is BB $\Rightarrow$ BC or oscillation. As is clear in Figure 12. if the universe starts from a big bang it expands to $a_{\rm{T}}$, then turns back at $a_{\rm{T}}$ and experiences a big crunch. If the universe exists initially in the region $[a_{\rm{min}},a_{\rm{max}}]$  with oscillating
behavior,   after some oscillations the quantum tunneling to  $a_{\rm{T}}$ may happen and the universe faces with a big crunch. The period of an oscillation can be calculated by means of $T=2\int_{a_{\rm{min}}}^{a_{\rm{max}}}\frac{da}{\sqrt{-V(a)}}$.

\begin{figure*}[ht]
  \centering
  \includegraphics[width=2.5in]{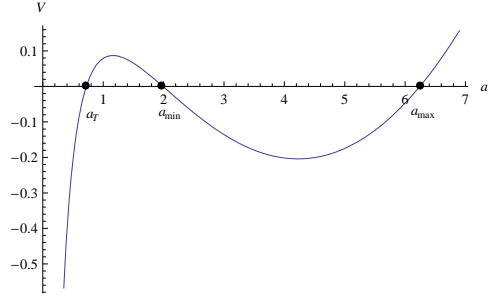}\hspace{2cm}
  \caption{The potential $V(a)$  with model parameters satisfying  $\frac{3\beta_{2}}{4}\left(1+9\beta_{2}\right)<\beta_{4}<\frac{9\beta_{2}^{2}}{\beta_{0}}$ shows an oscillating universe or a BB $\Rightarrow$ BC one with  the constants as $m=1$, $M_{g}=M_{f}$, $M_{*}=1$, $\beta_{0}=1$, $\beta_{2}=2.1$, $\beta_{4}=35$, $\lambda=1.2$ and with the radii  $a_{\rm{T}}=0.72092$, $a_{\rm{min}}=1.98901$ and $a_{\rm{max}}=6.2453$. The period of an oscillation is $T=29.7726$.}
  \label{stable11}
\end{figure*}

If $\lambda=\lambda^{+}$, $a_{\rm{T}}$ and $a_{\rm{min}}$ coincide with each other and give a double positive root $a_{\rm{s}}^{+}$ defined in (\ref{Fbi75}) shown in Figure 13. The point $a_{\rm{s}}^{+}$ implies an unstable ES solution and the universe can turn back at $a_{\rm{s}}^{+}$ or  $a_{\rm{max}}$ but the cosmic evolution type is BB $\Rightarrow$ BC.

\begin{figure*}[ht]
  \centering
  \includegraphics[width=2.5in]{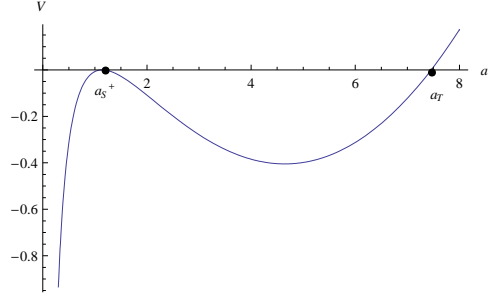}\hspace{2cm}
  \caption{The potential $V(a)$ under the condition $\lambda=\lambda^{+}$ with model parameters satisfying  $\frac{3\beta_{2}}{4}\left(1+9\beta_{2}\right)<\beta_{4}<\frac{9\beta_{2}^{2}}{\beta_{0}}$ turns out an unstable ES universe with the constants as $m=1$, $M_{g}=M_{f}$, $M_{*}=1$, $\beta_{0}=1$, $\beta_{2}=2.1$, $\beta_{4}=35$, $\lambda=1.30585$ and with the radii  $a_{\rm{s}}^{+}=1.14381$ and $a_{\rm{max}}=7.4591$.}
  \label{stable12}
\end{figure*}
Moreover, when $0<\lambda=\lambda^{-}$ we will see that $a_{\rm{min}}$ and $a_{\rm{max}}$ coincide with each other and we have a double solution $a_{\rm{s}}^{-}$. This is a stable ES solution and the cosmic evolution type is  BB $\Rightarrow$ BC because for $a\in (0,a_{\rm{T}}]$ and $a=a_{\rm{s}}^{-}$ the potential is $V(a)\leq 0$. As shown in Figure 14, there is a quantum tunneling possibility in the range $a\in [a_{\rm{T}},a_{\rm{s}}^{-}]$, so if the universe stays at $a=a_{\rm{s}}^{-}$ initially it may goes through the big crunch evolution. Conversely, if the universe starts from the big bang, a quantum tunneling may happen at $a_{\rm{T}}$ and the evolution ends up with a stable ES state, or after the expansion regime it turns over at  $a_{\rm{T}}$ and the big crunch regime starts.

\begin{figure*}[ht]
  \centering
  \includegraphics[width=2.5in]{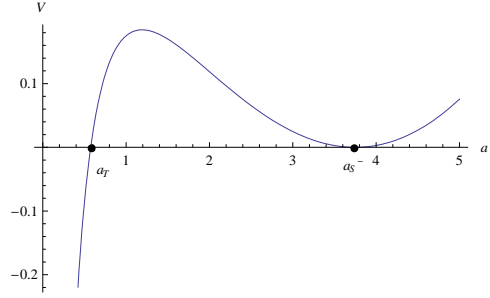}\hspace{2cm}
  \caption{The potential $V(a)$ under the condition $\lambda=\lambda^{-}$ with model parameters satisfying  $\frac{3\beta_{2}}{4}\left(1+9\beta_{2}\right)<\beta_{4}<\frac{9\beta_{2}^{2}}{\beta_{0}}$ leads to a stable ES universe with the constants as $m=1$, $M_{g}=M_{f}$, $M_{*}=1$, $\beta_{0}=1$, $\beta_{2}=2.1$, $\beta_{4}=35$, $\lambda=1.08159$ and with the radii  $a_{\rm{T}}=0.574419$ and $a_{\rm{s}}^{-}=3.74839$.}
  \label{stable13}
\end{figure*}

Eventually, we can consider a triplet root possibility in which $\lambda=\lambda^{-}=\lambda^{+}$. It means that $a_{\rm{T}}$,  $a_{\rm{min}}$ and  $a_{\rm{max}}$ coincide with each other. To obtain this, we should look back to (\ref{Fbi77}) which reads
as
\begin{align}\label{Fbi82}
\beta_{4}\rightarrow \left(\frac{3\beta_{2}M_*^2}{4}\left(1+9\beta_{2}M_*^2\right)\right)^{+},
\end{align}
which is applied in Figure 15.  showing a BB$\Rightarrow$BC universe.\\

\begin{figure*}[ht]
  \centering
  \includegraphics[width=2.5in]{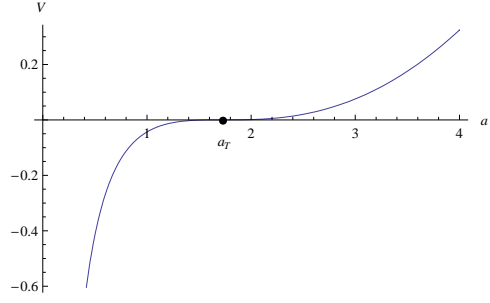}\hspace{2cm}
  \caption{The potential $V(a)$ under the condition $\lambda=\lambda^{-}=\lambda^{+}$ with model parameters satisfying  $\beta_{4}\rightarrow \left(\frac{3\beta_{2}M_*^2}{4}\left(1+9\beta_{2}M_*^2\right)\right)^{+}$ leads to a BB $\Longrightarrow$ BC universe with the constants as $m=1$, $M_{g}=M_{f}$, $M_{*}=1$, $\beta_{0}=1$, $\beta_{2}=2.1$, $\beta_{4}=31.34255$, $\lambda=1.3839$ and with the radii  $a_{\rm{T}}=1.76538$.}
  \label{stable14}
\end{figure*}

~~~~~~~~~~~~~~~~~~~~~~~~~~~~~~~~~~~~~~~~~~~~~~~~~~~~~~~~~~\textbf{2}. Two positive roots\\

Assuming $a_{1}<0$ and $a_{2}$, $a_{3}>0$, we  set $a_{2}=a_{\rm{min}}$ and $a_{3}=a_{\rm{max}}$. This classification will not be studied here since two positive roots  require that $a_{\rm{min}}a_{\rm{max}}a_{\rm{T}}$ becomes negative which leads to a negative $\lambda$.\\

Considering ``One positive root" and ``No positive root", again we have to ignore these parts because the relations (\ref{Fbi79}) to (\ref{Fbi81}) imply that a  positive $\lambda$s requires all these three terms to  be positive.\\

~~~~~~~~~~~~~~~~~~~~~~~~~~~~~~~~~~~~~~~~~~~~~~~~~~~~~~~~~~~~~~~~~~~\textbf{B}. $\lambda>\lambda^{+}$\\

The calculations here are similar to the previous ones, so we just bring the results. In Figure 16. we obtain a BB$\Rightarrow$BC universe while we still obey the inequality (\ref{Fbi80}).\\

\begin{figure*}[ht]
  \centering
  \includegraphics[width=2.5in]{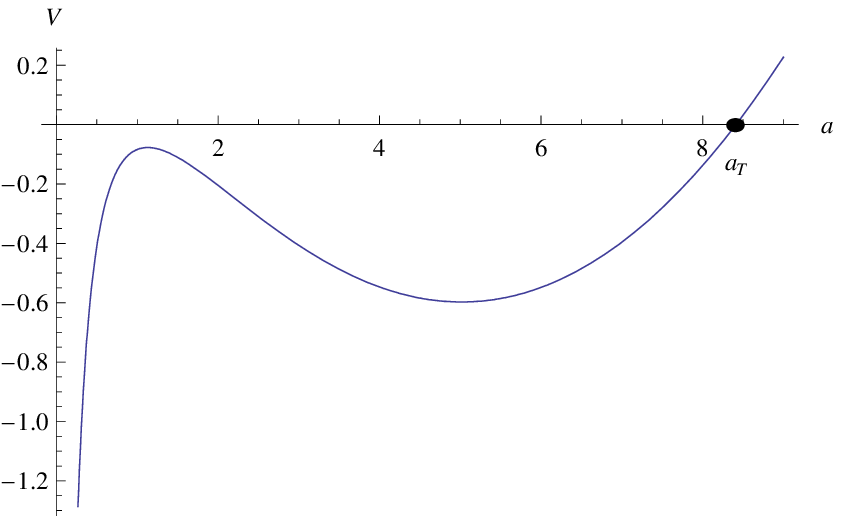}\hspace{2cm}
  \caption{The potential $V(a)$ under the condition $\lambda>\lambda^{+}$ with model parameters satisfying  $\frac{3\beta_{2}M_*^2}{4}\left(1+9\beta_{2}M_*^2\right)<\beta_{4}<\frac{9\beta_{2}^{2}}{\beta_{0}}$ turns out a BB $\Longrightarrow$ BC universe with the constants as $m=1$, $M_{g}=M_{f}$, $M_{*}=1$, $\beta_{0}=1$, $\beta_{2}=2.1$, $\beta_{4}=35$, $\lambda=1.3839$ and with the radii  $a_{\rm{T}}=1.41332$.}
  \label{stable15}
\end{figure*}

~~~~~~~~~~~~~~~~~~~~~~~~~~~~~~~~~~~~~~~~~~~~~~~~~~~~~~~~~~~~~~~~~\textbf{C}. $0<\lambda<\lambda^{-}$\\

Figure 17. again shows a BB$\Rightarrow$BC cosmic evolution type universe.\\

\begin{figure*}[ht]
  \centering
  \includegraphics[width=2.5in]{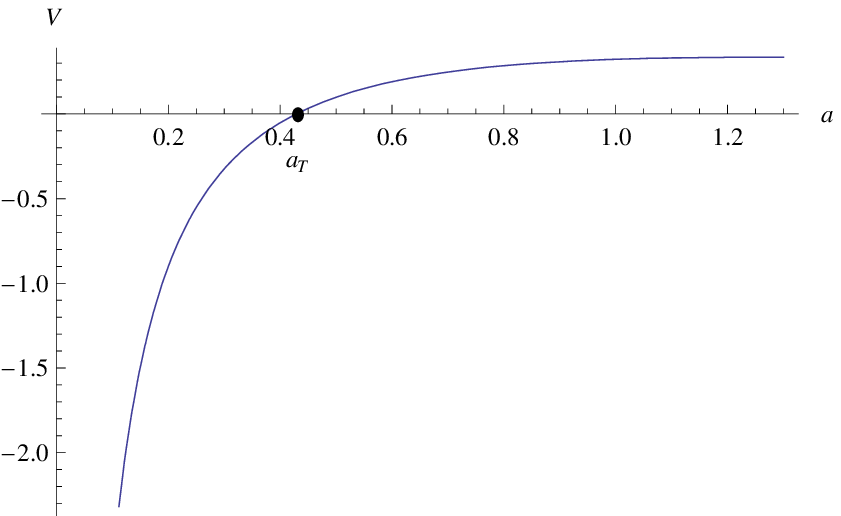}\hspace{2cm}
  \caption{The potential $V(a)$ under the condition $0<\lambda<\lambda^{-}$ with model parameters satisfying  $\frac{3\beta_{2}M_*^2}{4}\left(1+9\beta_{2}M_*^2\right)<\beta_{4}<\frac{9\beta_{2}^{2}}{\beta_{0}}$ turns out a BB $\Longrightarrow$ BC universe with the constants as $m=1$, $M_{g}=M_{f}$, $M_{*}=1$, $\beta_{0}=1$, $\beta_{2}=2.1$, $\beta_{4}=35$, $\lambda=0.9$ and with the radii  $a_{\rm{T}}=0.428839$.}
  \label{stable16}
\end{figure*}

\textbf{CaseII}: $\kappa=-1$ \\

In the open universe case, the potential is obtained as follows

\be\label{Fbi83}
V(a)=-1+m^{2}a^{2}\times\frac{\frac{-3\beta_{2}^{2}}{\beta_{4}}+\frac{\beta_{0}}{3}+\frac{\lambda}{3m^{2}M_{g}^{2}}\left(\frac{1}{a^{3}}+\frac{1}{a}\right)}{
\frac{3M_{*}^{2}\beta_{2}}{\beta_{4}}-1}.
\ee

Considering the conditions $V(a)=0$ and $V'(a)=0$, we extract the following ES solutions
\begin{eqnarray}\label{eq84}
\lambda^{\pm}&=&\frac{1}{4\sqrt{2}\beta_{4}\sqrt{m^{2}\left(9\beta_{2}^{2}-\beta_{0}\beta_{4}\right)}}M_{g}^{2}\left(27 m^2 \beta _2^2+9 M_*^2 \beta _2-3 \left(\beta _0 m^2+1\right) \beta _4\right.\nonumber\\&&\left.\pm\sqrt{3} \sqrt{\left(9 m^2 \beta _2^2-9 M_*^2 \beta _2+\left(3-m^2 \beta _0\right) \beta _4\right) \left(27 m^2 \beta _2^2-3 M_*^2 \beta _2+\left(1-3 m^2 \beta
   _0\right) \beta _4\right)}\right)\times \nonumber\\&& \left(-27 m^2 \beta _2^2+9 M_*^2 \beta _2+3 \left(m^2 \beta _0-1\right) \beta _4\right.\nonumber\\&&\left.\mp\sqrt{3} \sqrt{\left(9 m^2 \beta _2^2-9 M_*^2 \beta _2+\left(3-m^2 \beta _0\right) \beta _4\right) \left(27 m^2 \beta _2^2-3 M_*^2 \beta _2+\left(1-3 m^2 \beta
   _0\right) \beta _4\right)}\right)^{\frac{1}{2}}
\end{eqnarray}

\be\label{Fbi85}
a_{\rm{S}}^{\pm}=
\sqrt{\frac{9 \left(1-3 \beta _2\right) \beta _2+3 \left(\beta _0-1\right) \beta _4\mp\sqrt{3} \sqrt{\left(3 \beta _2 \left(9 \beta _2-1\right)+\left(1-3 \beta
   _0\right) \beta _4\right) \left(9 \left(\beta _2-1\right) \beta _2-\left(\beta _0-3\right) \beta _4\right)}}{18 \beta _2^2-2 \beta _0 \beta _4}}.
\ee

Again, since $a_{\rm{T}}^{\pm}$ is too large to be mentioned here, we just use its numerical value in the corresponding figures. Now, we find the following
allowed ranges in which the above quantities become real

\begin{align}\label{Fbi86}
\beta_{4}>\rm{Max} \left\{\frac{9\beta_{2}^{2}}{\beta_{0}},3\beta_{2}M_{*}^{2}\right\}.
\end{align}
It is worthwhile to mention that the requirement of having real $\lambda^{\pm}$ and $a_{\rm{S}}^{\pm}$ leads to  positive $a_{\rm{S}}^{\pm}$ with negative $\lambda^{\pm}$. However, we cannot obtain positive $a_{\rm{S}}^{\pm}$ with a positive $\lambda$, yet there is a possibility to have a positive $a_{\rm{T}}$ with a positive $\lambda$. As a result, the only possible solution is ``one positive root" leading to just one $\lambda>0$. Thus, we suppose the  positive $\lambda$ by which the positive $a_{\rm{T}}$ is obtained, as

\begin{eqnarray}\label{Fbi87}
\lambda&=&\sqrt{\frac{9 \left(1-3 \beta _2\right) \beta _2+3 \left(\beta _0-1\right) \beta _4+\sqrt{3} \sqrt{\left(3 \beta _2 \left(9 \beta _2-1\right)+\left(1-3 \beta
   _0\right) \beta _4\right) \left(9 \left(\beta _2-1\right) \beta _2-\left(\beta _0-3\right) \beta _4\right)}}{18 \beta _2^2-2 \beta _0 \beta _4}}\times  \\\nonumber&& \frac{ \left(-9 \beta
   _2 \left(3 \beta _2+1\right)+3 \left(\beta _0+1\right) \beta _4+\sqrt{3} \sqrt{\left(3 \beta _2 \left(9 \beta _2-1\right)+\left(1-3 \beta _0\right) \beta
   _4\right) \left(9 \left(\beta _2-1\right) \beta _2-\left(\beta _0-3\right) \beta _4\right)}\right)}{4 \beta _4}.
\end{eqnarray}

In order to have a real and positive $\lambda$, we should obey the following inequality

\begin{align}\label{Fbi88}
3\beta_{2}M_{*}^{2}<\beta_{4}<\frac{9\beta_{2}^{2}}{\beta_{0}}.
\end{align}

Thus, by using the above terms, we plot Figure 18. showing a BB$\Rightarrow$BC universe.\\

\begin{figure*}[ht]
  \centering
  \includegraphics[width=2.5in]{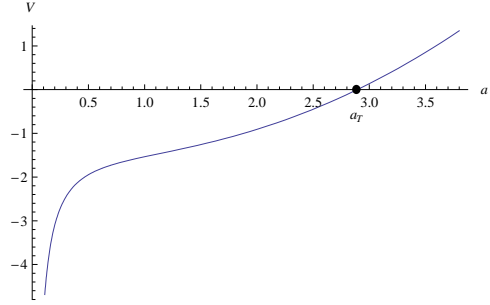}\hspace{2cm}
  \caption{The potential $V(a)$ under the condition $0<\lambda$ with model parameters satisfying  $3\beta_{2}M_{*}^{2}<\beta_{4}<\frac{9\beta_{2}^{2}}{\beta_{0}}$ turns out a BB $\Longrightarrow$ BC universe with the constants as $m=1$, $M_{g}=M_{f}$, $M_{*}=1$, $\beta_{0}=1$, $\beta_{2}=3.5$, $\beta_{4}=65$, $\lambda=1.02172$ and with the radii  $a_{\rm{T}}=2.89235$.}
  \label{stable17}
\end{figure*}

\section{Conclusions\label{Sec6}}
Massive bigravity is a modification of general relativity and also massive gravity. The idea behind massive bigravity in which we have two dynamical metric tensors improves the cosmology of the theory with two additional modified first and second Friedmann equations. Using two modified first Friedmann equations of $g_{\mu\nu}$ and $f_{\mu\nu}$ respectively, we have investigated all possible cosmic evolutions with a method in which the dynamics of the scale factor (the scale factor of metric $g_{\mu\nu}$ which is coupled to source) behaves similar to that of a particle moving under  a potential. The potential contains the mass term constructed by two metrics respecting the symmetries of spatial isotropy and homogeneity, with the energy density which is coupled to the metric $g_{\mu\nu}$. Considering three energy density classifications, matter-dominated, radiation-dominated universes and universe with dust and phantom, we find different kinds of cosmic evolutions of the early universe in the context of massive bigravity. For the matter-dominated case in a closed universe we have extracted a big bang to big crunch evolution or a bouncing universe, an unstable Einstein static universe and a big bang to infinity or infinity to big crunch, while for an open universe we just find a BB $\Longrightarrow$ BC one.  Finally for a spatially flat case, we cannot find any analytical solution for $a$ and $\lambda$ in terms of the free parameters $\beta_{i}$'s, nevertheless we find a bouncing universe,  numerically. In the radiation-dominated regime, only for a closed universe we have obtained a universe which starts from a big bang,  expands to an unstable Einstein static universe and may then bounce to a big crunch.   Eventually, in this context we have studied spatially flat universe numerically without extracting any explicit  expression for $a_{\rm{T}}$ and $\lambda$ which have led to a big bang to big crunch evolution.

 The main subsection belongs to the  ``universe with dust and phantom" in which we have found an oscillating universe which is considered as a model implying that the early universe was oscillating between two scale factors $a_{\rm{min}}$ and $a_{\rm{max}}$, past eternally. Moreover, in our model the oscillation may end up with a big crunch via a quantum tunneling which means that if the universe exists in the region [$a_{\rm{min}},a_{\rm{max}}$] initially,  one  or several oscillations may happen and afterwards the quantum tunneling occurs to $a_{\rm{T}}$  and then it experiences a big crunch evolution. Another possibility in this regime is obtained when the big bang happens and expands to an unstable ES state in which the potential equals to zero at $a_{\rm{S}}^{+}$ and then because of the instability the expansion starts again towards $a_{\rm{T}}$ point at which it turns back to big crunch, or the universe may turn back at $a_{\rm{S}}^{+}$ and starts the big crunch evolution. In another circumstance, our universe can stay in a stable ES state past eternally and then experiences
a quantum tunnelling to a big crunch regime, or it may start from the big bang initially and expand and then experiences
a quantum tunnelling to a stable ES state  and eventually for this closed case we find a BB $\Longrightarrow$ BC universe.

 In the open universe case, we just have bounce evolution and  bouncing universe with deferent bouncing points. Actually, our universe can exist in one of bouncing periods which means that the big bang resulted of its previous big crunch happened and
then it is experiencing accelerated expansion which may turn back at a bouncing scale factor and again evolves to a big crunch and eventually we studied spatially flat universe numerically without extracting any explicit expression for $a_{\rm{T}}$ and $\lambda$ which led to a big bang to big crunch evolution. Additionally, refereing to the obtained terms for $\lambda^{\pm}$ and $a_{\rm{S}}^{\pm}$ in all sections, we realize that the roll of   increasing or  decreasing graviton mass $m$ means that if the graviton mass becomes large enough at early universe then it gives rise to a very small scale factor and also a very small $\lambda$ which are desirable for our theory. As we know, $\lambda=\rho_{0}a_{0}^{3}$ contains the present day energy density and the present day scale factor which becomes small-valued when the graviton mass becomes large. Moreover, the ES scale factor $a_{\rm{S}}$ becomes as small as possible when the graviton mass goes to a very large value $m\sim 10^{12}$$\rm{GeV}$. In conclusion, this paper is devoted to the study of early universe with different kinds of evolutions  in massive bigravity, which is influenced by the mass term of this theory by which the leading term of the scale factor in the potential plays an important role. Also we should note that the sub-model of massive bigravity with $\beta_{1}=\beta_{3}=0$, called the minimal massive bigravity model, covers the early universe evolutions in the general relativity theory.

\section{Acknowledgments}
This work has been supported financially by Iran National Science Foundation (INSF) under postdoctoral research project No. 96007730.

%\appendix

\end{document}